\documentclass[fleqn,usenatbib]{mnras}

\usepackage{newtxtext,newtxmath}
\usepackage[T1]{fontenc}
\usepackage{ae,aecompl}

\usepackage{graphicx}	% Including figure files

\usepackage[compatibility=false]{caption}
\usepackage{subcaption}

\defcitealias{Wang_2017}{WG17}

\title[X-ray driving of winds]{The importance of X-ray frequency in driving photoevaporative winds}

\author[A. D. Sellek et al.]{
Andrew D. Sellek,$^{1}$\thanks{E-mail: ads79@cam.ac.uk}
Cathie J. Clarke,$^{1}$
Barbara Ercolano,$^{2,3}$
\\
$^{1}$Institute of Astronomy, University of Cambridge, Madingley Road, Cambridge CB3 0HA, UK\\
$^{2}$Universitats-Sternwarte, Ludwig-Maximilians-Universit\"{a}t M\"{u}nchen, Scheinerstr. 1, D-81679 M\"{u}unchen, Germany\\
$^{3}$Excellence Cluster Origins, Boltzmannstrasse 2, D-85748 Garching bei M\"{u}nchen, Germany\\
}

\date{Accepted XXXX. Received YYYY; in original form ZZZZ}

\pubyear{2021}

\begin{document}
\label{firstpage}
\pagerange{\pageref{firstpage}--\pageref{lastpage}}
\maketitle

\begin{abstract}
Photoevaporative winds are a promising mechanism for dispersing protoplanetary discs, but so far theoretical models have been unable to agree on the relative roles that the X-ray, Extreme Ultraviolet or Far Ultraviolet play in driving the winds. This has been attributed to a variety of methodological differences between studies, including their approach to radiative transfer and thermal balance, the choice of irradiating spectrum employed, and the processes available to cool the gas. We use the \textsc{mocassin} radiative transfer code to simulate wind heating for a variety of spectra on a static density grid taken from simulations of an EUV-driven wind. We explore the impact of choosing a single representative X-ray frequency on their ability to drive a wind by measuring the maximum heated column as a function of photon energy. We demonstrate that for reasonable luminosities and spectra, the most effective energies are at a few $100~\mathrm{eV}$, firmly in the softer regions of the X-ray spectrum, while X-rays with energies $\sim1000~\mathrm{eV}$ interact too weakly with disc gas to provide sufficient heating to drive a wind. We develop a simple model to explain these findings. We argue that further increases in the cooling above our models - for example due to molecular rovibrational lines - may further restrict the heating to the softer energies but are unlikely to prevent X-ray heated winds from launching entirely; increasing the X-ray luminosity has the opposite effect. The various results of photoevaporative wind models should therefore be understood in terms of the choice of irradiating spectrum.
\end{abstract}

\begin{keywords}
circumstellar matter -- protoplanetary discs -- radiative transfer -- X-rays: stars
%hydrodynamics -- accretion, accretion discs --   -- submillimetre: planetary systems -- planets and satellites: formation
\end{keywords}

%%%%%%%%%%%%%%%%%%%%%%%%%%%%%%%%%%%%%%%%%%%%%%%%%%

%%%%%%%%%%%%%%%%% BODY OF PAPER %%%%%%%%%%%%%%%%%%

\section{Introduction}
Though first proposed to explain the long lifetime of \ion{H}{II} regions around massive stars by resupplying them with material \citep[e.g.][]{Hollenbach_1994}, photoevaporative winds are now one of the most-promising mechanisms for dispersing protoplanetary discs \citep{EP_2017,Kunitomo_2020} and thus ending the era of planet formation around a young star. The major successes of photoevaporative models include clearing discs from inside-out \citep{Koepferl_2013}; the production of the so-called "two-timescale" behaviour \citep[as opposed to the gradual power law decline of a purely viscous model,][]{Hartmann_1998} in which this dispersal is rapid due to the action of the UV-switch \citep{Clarke_2001}, though some relic discs may remain \citep{Owen_2011,Owen_2012}; and reproducing the dependence of inner disc lifetime on stellar mass \citep{Komaki_2021,Picogna_2021}.

Photoevaporative winds are thermally launched: they occur when the upper layers of a disc become heated by high energy radiation from either the central star (\textit{internal} photoevaporation) or a massive neighbour (\textit{external} photoevaporation) and the resulting pressure gradients drive material off of the disc. Broadly speaking, this is expected to happen for radii $r \gtrsim \frac{GM_*}{c_S^2}$ (for stellar mass $M_*$ and sound speed in the wind $c_S$) where the available thermal energy is enough to overcome the gravitational potential of the star \citep{Shu_1993,Hollenbach_1994,Font_2004,Alexander_2014,Clarke_2016,Sellek_2021}. Equivalently, one can invert this to say that at a radius $R$, material must be heated above the escape temperature $T_{\rm esc} \sim \frac{GM_* \mu m_H}{2 k_B R}$ \citep{Owen_2012}.

In order to explain the ionised surroundings of massive stars, it was consequently natural for the earliest theories \citep{Hollenbach_1994} to assume the heating was due to the Extreme Ultraviolet (EUV) at energies $\gtrsim 13.6~\mathrm{eV}$. % which leads to regions of complete ionisation that are ionisation bounded.
Though most stars are of a lower mass with lower temperature and UV fluxes, the derived mass loss profile scaled with ionising photon count ($\Phi$) and so could be applied to solar mass stars by using a lower $\Phi$ value \citep{Shu_1993}.
However, without a strong contribution from the photospheric blackbody spectrum, the origin and magnitude of such an ionising flux for low mass stars is more debated and observationally challenging to determine due to absorption by hydrogen in the line of sight. \citet{Alexander_2004a} ruled out accretion hotpots as unable to produce a sufficient level of ionising photons to drive a significant wind, while possible chromospheric activity is poorly understood. Nevertheless, the value of $\Phi$ can exceed solar levels \citep[e.g.][]{Gahm_1979,Alexander_2005} with $10^{41}-10^{42}~\mathrm{s^{-1}}$ often assumed.

Ionising radiation also extends into the X-ray, which is much easier to constrain in surveys \citep[e.g.][]{Preibisch_2005,Gudel_2007}. As a consequence, the relationship between X-ray luminosity $L_X$, spectrum shape and stellar properties is better understood. Its impact on the heating of winds was first assessed by \citet{Alexander_2004b} who concluded the mass loss profiles were at best comparable to the EUV. However, X-rays drive winds from a larger area of the disc, potentially resulting in higher integrated mass-loss rates \citep{Ercolano_2009}. Moreover, these winds were definitively X-ray driven: attenuating the EUV part of \citet{Ercolano_2009}'s spectrum did not reduce their mass-loss rates since EUV photons cannot penetrate far into X-ray driven winds \citep{Ercolano_2010,Owen_2012} which are typically denser (but slower) and more neutral than their EUV equivalents.
They concluded softer X-rays $<1000~\mathrm{eV}$ were particularly important as once they pre-screened their spectrum enough to absorb these, a significant wind could no longer be launched.

The important role for X-ray suggested by static models was corroborated by the hydrodynamical calculations of \citet{Owen_2010,Owen_2011,Owen_2012}, in which most of the material never even passes into the EUV heated region.
%In these models, the wind is too dense and neutral for EUV to penetrate deeply, consequently it is of no consequence to driving the wind (most of which never passes through the EUV IF).
To make this calculation tractable, these works assumed thermal equilibrium with the temperatures prescribed as a pre-calculated function of density $n$ and local X-ray flux $L_X/r^2$ via the ionisation parameter $\xi = \frac{L_X}{nr^2}$ \citep{Tarter_1969}. This equilibrium relationship was established using the \textsc{mocassin} Monte Carlo radiative transfer code \citep{Ercolano_2003,Ercolano_2005,Ercolano_2008}.
The same methods, but with updated prescriptions that are also functions of the column density and use luminosity-dependent spectra, have been applied by \citet{Picogna_2019,Ercolano_2021,Picogna_2021} with qualitatively similar results.

On the other hand, the work of \citet[][hereafter WG17]{Wang_2017}, which aimed to better understand the line spectra of the winds by including thermochemistry in the model, suggests EUV has the dominant role.
The chemistry was handled using a simple chemical network of 24 species, with abundances updated according to reaction rates with each hydrodynamical timestep. Likewise to avoid assumption of thermal equilibrium they directly calculated heating rates from ray tracing - for simplicity using just 4 bins spanning the FUV, EUV and X-ray - and cooling rates from a variety of molecular and atomic processes.
These processes lead to a hotter, more tenuous, highly ionised, wind in which EUV photoionisation and adiabatic cooling were the key elements of the thermal balance, suggesting that thermal equilibrium cannot be assumed. Moreover, the X-rays were seemingly important only for helping puff up the underlying disc with molecular cooling processes responsible for offsetting their heating. The FUV was likewise important to the heating in these underlying layers. The overall mass loss rates were below those of X-ray models \citep[e.g.][]{Owen_2012}.

However, \citet{Nakatani_2018b} performed a similar exercise but found instead that at solar metallicity thermal winds were mostly FUV-driven, with X-rays  assisting mainly by increasing ionisation levels, thus allowing FUV to penetrate more deeply. In the absence of FUV, X-rays did not drive a wind in these models, regardless of the hardness of the X-ray spectrum.
%- but cannot drive a wind in the absence of FUV regardless of spectral hardness.

The thermal state of the wind is important for understanding wind observable quantities such as line spectra \citep[e.g.][]{Font_2004,Ercolano_2016,Ballabio_2020,Weber_2020}.
For example, an X-ray wind would mean slower, cooler winds than in the EUV case, which manifests in the centroid shift and full width at half maximum (FWHM) of collisionally excited lines (which are both functions of the sound speed in the wind); this is seen for the \ion{O}{I} $6300~\text{\r{A}}$ line \citep{Ballabio_2020}.
Since it is produced by a neutral species, such a line requires a relatively low level of wind ionisation, which is not very consistent with an EUV model \citep{Font_2004} meaning that the observed lines have a more natural explanation if the wind is X-ray driven. \citet{Ercolano_2016} showed how this line could be produced by an X-ray driven wind, although since such high temperatures are required for its production its luminosity nevertheless scales with the EUV luminosity.
Moreover, the mass loss rates have a key impact on demographic indicators of disc dispersal such as the $\dot{M}_{\rm acc}-M_{\rm disc}$ plane \citep{Somigliana_2020,Sellek_2020b}, the properties of transition discs \citep[e.g.][]{Owen_2011,Owen_2012,Picogna_2019,Ercolano_2021} and the correlation of accretion rates and dispersal times with stellar properties \citep{Ercolano_2014,Flaischlen_2021,Picogna_2021}.
Therefore it is important that we resolve the tensions between the different studies outlined above with respect to the most important heating and cooling mechanisms if we are to establish accurate predictions of observables from photoevaporative models.

We can summarise five key differences: a) the treatment of radiative transfer, b) the shape and resolution of the irradiating spectrum c) the atomic cooling processes included, d) the inclusion of molecules and molecular cooling and e) the assumption (or lack thereof) of thermochemical equilibrium.
Here we wish to address the question of driving radiation by exploring the first three of these through irradiating the \citetalias{Wang_2017} density grids using \textsc{mocassin} setup to create the conditions in their simulations. Specifically this allows us to comment on the consistency between the Monte Carlo and ray tracing radiative transfer methods, vary the spectral bins to establish the efficacy of different X-ray bands, and estimate the contributions of different mechanisms to thermochemical balance.

%First we discuss in more detail the differences between the approaches and our modelling choices to create a similar setup within \text{mocassin} in Section 2.
%In Section 3, we then demonstrate to what extent this modelling can achieve a temperature structure that is reasonably consistent with \citetalias{Wang_2017} and how our modelling choices impact different regions of the disc and wind.
The aim of this paper is to explore how the role of X-rays in driving photoevaporative winds depends on the spectrum of the X-rays employed. In order to investigate this issue we will first attempt to reproduce the results of \citetalias{Wang_2017} using the X-ray spectrum employed in their work (i.e. $1000~\mathrm{eV}$ X-rays).
Following a description of our radiative transfer modelling framework and its differences from previous work (Section 2), we describe in Section 3 how choices in the treatment of optical line cooling impact different regions of the disc and wind and find those required in order to produce a similar temperature structure to \citetalias{Wang_2017}.
This then equips us to examine in Section 4 how the temperature structure is modified when we change the X-ray spectrum. We show that a modest (factor 2) reduction in the characteristic frequency of the X-ray results in a substantial (order of magnitude) increase in the column that can be heated by X-rays and that such a change inverts the conclusion of \citetalias{Wang_2017} with regard to the relative importance of UV and X-rays in driving thermal disc winds. We also show how the result that X-ray heating is optimised at energies around 500 eV can be quantitatively well-understood in terms of a simple analytic model of radiative balance between cooling processes and X-ray heating.
%This leaves us in the position to explore the role of X-ray frequency in Section 4, where we show results for a variety of spectra and demonstrate that softer X-ray energies can indeed launch a wind, considering their relative abilities to do so in the context of a simplified model of radiative balance between cooling processes and X-ray heating.
In Section 5 we discuss the contributions of different mechanisms to thermochemical balance in order to critically assess the need to include molecular cooling and the validity of the equilibrium assumption.
Finally we present our conclusions and discuss the impact they might have on disc evolution in Section 6.

\section{Model description}
\subsection{Key differences between previous works}
The methodology used by different photoevaporation studies has varied substantially: \citetalias{Wang_2017} \citep[and similarly][]{Nakatani_2018b} conducted radial ray tracing to calculate an attenuated flux in each cell in their simulations using the optical depth provided by a range of photoreactions. This process ignores the potential for scattering of radiation, the diffuse EUV field produced by recombinations, and the ability of these processes to change the frequency of radiation. The diffuse EUV has typically been thought important to photoevaporation \citep{Hollenbach_1994} although this is somewhat debated \citep{Tanaka_2013,Hollenbach_2017} and may depend on any assumed disc structure. However, it provides a relatively inexpensive way of estimating the radiation field, allowing them to avoid assuming thermal equilibrium but instead update the ionisation states of material and perform photoionisation heating after each hydrodynamical timestep. That is to say, the thermal evolution of the disc/wind material is calculated by operator splitting, with atomic and molecular heating/cooling in one substep, and the hydrodynamical terms  - adiabatic cooling by PdV work and advection of thermal energy (hereafter collectively hydrodynamical cooling) - in the other.

On the other hand, most of the works favouring X-ray photoevaporation \citep{Owen_2010,Owen_2011,Owen_2012,Picogna_2019,Ercolano_2021,Picogna_2021} have been based on calculations using \textsc{mocassin} \citep{Ercolano_2003,Ercolano_2005,Ercolano_2008}, a Monte Carlo code which releases a number of packets of fixed energy into a fixed density grid at frequencies randomly sampled from the input spectrum. As the energy packets pass through cells they have an absorption probability; if absorbed the packet is re-emitted in a random direction at a frequency randomly sampled from the local emissivity. This process inherently conserves energy in the radiation packets and so it requires an assumption of thermal equilibrium between radiative heating and cooling processes - once the packets have passed through the grid, the radiative intensity can be estimated and the temperature and ionisation in each cell are updated iteratively until they give local heating and cooling rates that are equal. This procedure is then iterated until the solution converges.

Since this process is computationally more expensive, it would be impractical to run this at each timestep of a hydrodynamical simulation.
Instead, assuming X-ray dominates the heating, the usual procedure is to use precalculate a relationship between the temperature, density and X-ray flux of simulation cells via the ionisation parameter $\xi$. This can then be used to calculate the temperatures in a hydrodynamical simulation at each step without performing radiative transfer each time. While originally the fluxes were based only on geometric dilution \citep{Owen_2010}, attenuation can be accounted for by providing fits at a range of column densities \citep{Picogna_2019}. Post-processing with \textsc{mocassin} can then used to confirm the self-consistency of the resulting solution \citep{Owen_2010,Picogna_2019}. Although \textsc{mocassin} has been benchmarked against known solutions for the thermal structure of various setups, this does not guarantee its accuracy for this particular problem, particularly given that it necessarily ignores hydrodynamic contributions to the thermal balance.

Another major methodological difference between previous works is the irradiating spectrum. The \citet{Ercolano_2009} spectrum used by works from \citet{Owen_2010} to \citet{Picogna_2019} covers energies from $11.27~\mathrm{eV}$ (i.e. just above the first ionisation energy of C) to around $12~\mathrm{keV}$ and is sampled by \textsc{mocassin} to 1396 different energy bins. The spectrum is based on the coronal emission of RS CVn active binaries and has roughly similar luminosities in the EUV ($13.6-100~\mathrm{eV}$) and X-ray ($>100~\mathrm{eV}$) bands.

Whereas, \citetalias{Wang_2017} model the spectrum using just 4 bands at $7~\mathrm{eV}$, $12~\mathrm{eV}$, $25~\mathrm{eV}$ and $1000~\mathrm{eV}$. Their spectrum, \citep[which follows][]{Gorti_2009a} is overall softer than that of \citet{Ercolano_2009} in that it contains $6.25$ times more energy in the EUV band relative to the X-ray, and the representative X-ray is also slightly softer than the average of the \citet{Ercolano_2009} spectrum ($\int E_\nu H_\nu d\nu/\int H_\nu d\nu = 1086~\mathrm{eV}$). However, this attempt to represent energies covering two orders of magnitude perhaps does not really cover the extremes (particularly the softer end) of the X-ray band.
This begs the questions of both whether a) using just the four bands to spanning the whole spectrum are sufficient to resolve the true behaviour and b) if so whether these are a sensible choice.
We note that \citet{Nakatani_2018b} adopt an even softer spectrum where the X-ray luminosity is a further factor $\sim3$ lower compared to the EUV, and there is a considerably larger FUV luminosity.
%Although their spectrum is overall softer than that of \citet{Ercolano_2009} in that it contains nearly an order of magnitude more energy in the EUV band, the representative X-ray is somewhat harder than the average energy per X-ray photon of the \citet{Ercolano_2009} spectrum ($480~\mathrm{eV}$).
%Moreover, the photoionisation cross-section peaks at the ionisation energy of hydrogen ($13.6~\mathrm{eV}$) and is a steeply declining function of energy, making their choice of EUV band somewhat more penetrating than it might otherwise be.

Broadly speaking, both \textsc{mocassin} and the simulations of \citetalias{Wang_2017} contain a somewhat similar set of cooling processes; with the latter \citep[and also][]{Nakatani_2018b} also including some molecular cooling \citep[from collisionally-excited rovibrational states of $\mathrm{H_2}$, $\mathrm{H_2O}$, OH and CO, following][]{Neufeld_1993} and hydrodynamical cooling (as discussed above). This common set includes cooling by collisionally-excited Lyman $\alpha$ radiation from neutral hydrogen\footnote{\textsc{mocassin} also treats Lyman $\beta$ but this is everywhere subdominant to $\alpha$.}, collisionally-excited forbidden lines (CELs) of metal species, and dust-gas thermal accommodation.

However, the set of CELs used by \textsc{mocassin} is by far the more extensive as it is based on data from CHIANTI \citep{Dere_1997,Landi_2006} (though as noted by \citetalias{Wang_2017}, a few species including neutral sulfur are missing from its database), whereas \citetalias{Wang_2017} include only a select few following \cite{Tielens_1985}; these are all relatively low energy transitions in the IR\footnote{While \citetalias{Wang_2017} included \ion{O}{I} $6300~\text{\r{A}}$ as a coolant, they only modelled the decay of photodissociated OH, rather than collisional excitation.} and so are excited at fairly low temperatures, making them the dominant coolants at the modest temperatures of a photodissociation region (PDR), which is an appropriate description of the underlying disc which is mostly penetrated only by FUV \citep{Gorti_2008}.
However, in hot, (partially) ionised regions at or above the wind base, a number of different lines in the optical may come into play, such as the [\ion{O}{I}] $6300~\text{\r{A}}$ \& $6365~\text{\r{A}}$ and the [\ion{S}{II}] $4068~\text{\r{A}}$ \& $6718~\text{\r{A}}$/$6731~\text{\r{A}}$ doublet, many of which are important observational tracers of outflows from discs \citep{Hartigan_1995,Simon_2016,Fang_2018,Banzatti_2019}. Being higher energy transitions, these lines have higher excitation temperatures and cannot be excited in the colder disc, but become important once material reaches a few $1000~\mathrm{K}$: the optical lines tend to have both higher Einstein Coefficients, as well as imparting more energy per decay and so once excited can be highly effective coolants.

Moreover, \citetalias{Wang_2017} include an escape probability treatment based on the optical depth of their lines following \citet{Kwan_1981}. This is important if one wants to correctly estimate the line centre flux of any lines which are optically thick, such as resonance lines; practically speaking, this is only of consequence here for the Lyman lines which are permitted transitions and so have a higher oscillator strength than the forbidden lines by several orders of magnitude. \citetalias{Wang_2017} reduce their cooling rates in proportion to the escape probability, assuming that reabsorbed photons reheat the gas and do not cool it. However to reheat the gas, a collisional de-excitation of the excited state formed by photon reabsorption would be required, and the densities are everywhere orders of magnitude below the critical density of the Lyman $\alpha$ transition ($\sim3\times10^{12}~\mathrm{cm^{-3}}$) so collisional destruction of Lyman $\alpha$ is rare \citep{Dijkstra_2017}. More realistically the Lyman radiation should still escape, either in the optically thin line wings \citep[via a double diffusive process whereby reabsorbed photons perform a random walk in frequency e.g.][]{Avery_1968,Dijkstra_2017} or by being absorbed by dust and re-emitted at longer, optically-thin, wavelengths \citep{Cohen_1984} and thus the cooling rate matches the optically thin case \citep{Hollenbach_1979}\footnote{In reality, this is achieved at low densities by the cooling rate per excited atom being suppressed proportional to the escape probability but the excited population being increased in number in inverse proportion to the escape probability, resulting in no net effect; \citetalias{Wang_2017} do not, however, suggest whether reabsorption affects their calculation of the coolant density.}. \textsc{mocassin} therefore assumes that either way, this should not, ultimately, impede the cooling and does not reduce the cooling rates i.e. assumes the line is effectively optically thin.
This means a key difference between the methods is that Lyman cooling is likely several orders of magnitude more effective in \textsc{mocassin} than assumed by \citetalias{Wang_2017}.

A final difference to note is that the density grid used by \citetalias{Wang_2017} uses a larger inner boundary of $2~\mathrm{au}$ compared to the works favouring X-ray photoevaporation which use $0.33~\mathrm{au}$. This could potentially be important if there is significant attenuation of the EUV at $\lesssim 2~\mathrm{au}$. \citet{Nakatani_2018b}, who use an inner boundary of $1~\mathrm{au}$, report that varying their inner boundary to as little as $0.1~\mathrm{au}$ made little difference to the heating and ionization rates in the outer disc as there was not sufficient shielding by the inner regions of the disc and its atmosphere. However, while this is therefore unlikely to drive a difference between previous works, attenuation by material closer to the star than the inner boundary, for example accretion columns, remains possible and could affect the spectrum irradiating the disc and wind \citep[e.g.][]{Alexander_2004a}.

\subsection{Our models}
We aim to carry out radiative transfer in \textsc{mocassin} with conditions designed to replicate the approach of \citetalias{Wang_2017} within its existing framework. In this section we set out the details of how we achieve this.
Our simulations present are labelled in the form K\_Cooling %\_GH08\_Car90
representing the combination of a certain spectrum `K' with a certain cooling model.
The temperatures are completely calculated by \textsc{mocassin} using its iterative proceedure i.e. unlike some previous works \citep[e.g.][]{Owen_2010}, we do not fix them to dust temperatures from \citet{DAlessio_2001} at high column densities. Each model was run for eight iterations with $10^9$ photons and a final ninth iteration with $10^{10}$ photons, at which point they has all converged.

\begin{figure}
    \centering
    \includegraphics[width=\linewidth]{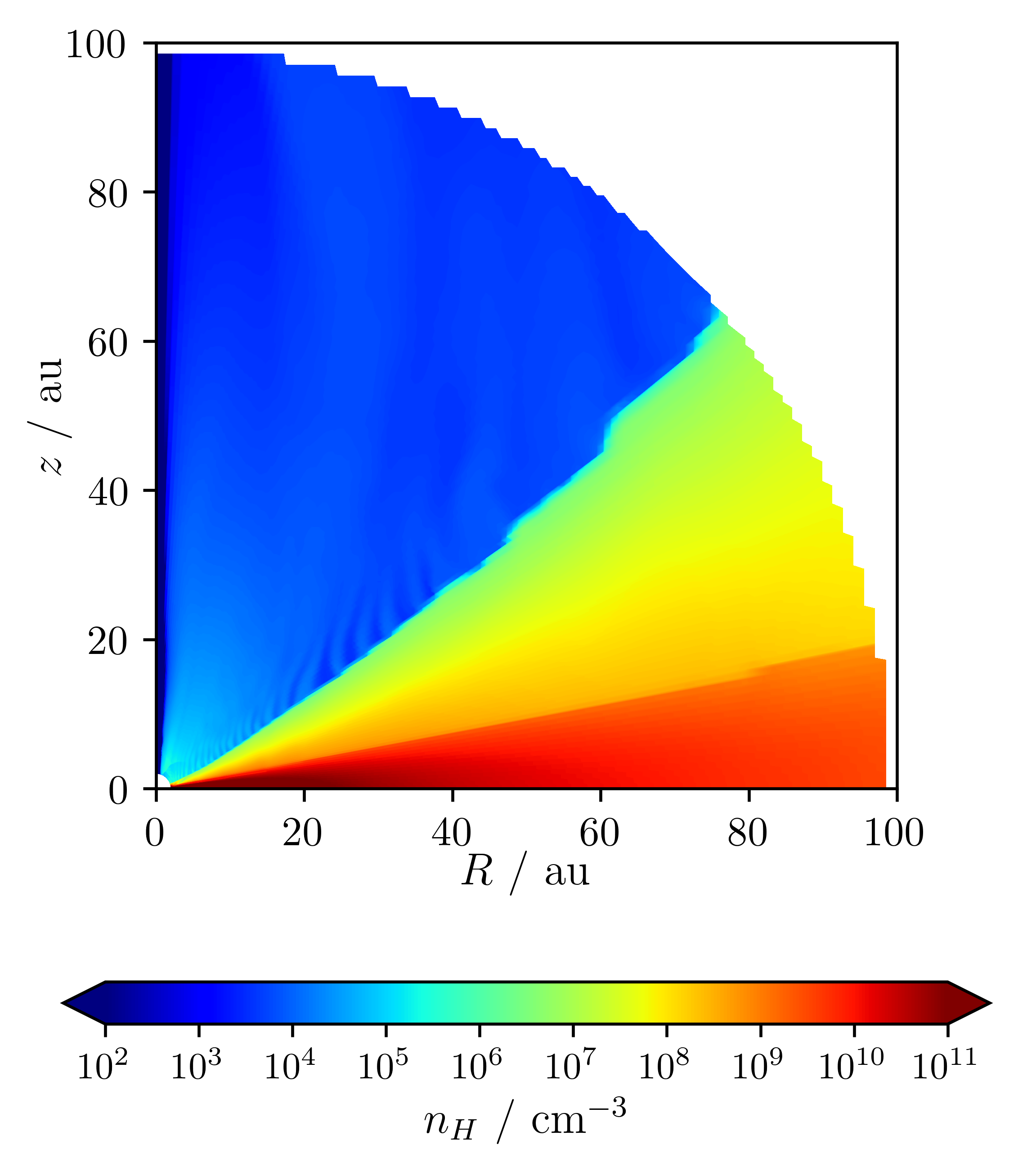}
    \caption{The interpolated density profile from \citetalias{Wang_2017} used for our calculations with \textsc{mocassin}.}
    \label{fig:dens}
\end{figure}

For each model we irradiate the same density profile for the wind and underlying disc; to aid comparisons to \citetalias{Wang_2017} we used the profile derived from their fiducial simulation. Since \textsc{mocassin} only accepts Cartesian grids, we interpolated onto a non-uniform Cartesian grid \citepalias[designed to provide more resolution at smaller radii, as with the logarithmic grid of][]{Wang_2017}. The total grid is $321\times321$ cells, of which 97724, with radii spanning $r=2-100~\mathrm{au}$, are active. The density profile is shown in Fig. \ref{fig:dens}.

\subsubsection{Spectra}
For our fiducial model, we replicate the spectrum of \citetalias{Wang_2017}. We set the nearest frequency bin in \textsc{mocassin} to have the same luminosity and all other bins to be zero. Note though that unlike in \citetalias{Wang_2017}, since the energy of the packets must be conserved, after interactions with atoms or dust the energy can be re-emitted in a different frequency bin so there will be secondary radiation at other energies. Since it follows \citetalias{Wang_2017} we label this spectrum with the key \textbf{W}; its luminosity in each band is listed in Table \ref{tab:spectra}. We note in particular that the X-ray luminosity $L_X$ is similar to the median value from surveys of T Tauri stars \citep[e.g.][]{Preibisch_2005,Gudel_2007}.

In Section \ref{sec:varybands}, we also consider a number of other spectra, including one with no X-ray and only UV (\textbf{U}), several with softer X-ray energy (\textbf{S\#\#\#} where \#\#\# is the energy in eV), and the \citet{Ercolano_2009} spectrum FS0H2Lx1 (\textbf{E}), which is a continuous spectrum as opposed to the rest which were all discrete.
The soft X-ray spectra have X-ray energies from $100~\mathrm{eV}$ to  $900~\mathrm{eV}$ in steps of $100~\mathrm{eV}$.
We normalise spectrum E to have the same EUV luminosity to control for the location of the ionisation front; as a result it has higher FUV and X-ray luminosities than the other spectra. To allow us to isolate the effects of increased X-ray luminosity from X-ray spectral shape, we thus also consider discrete spectra with the X-ray luminosity enhanced by a factor $6.25$ to match that of spectrum E (\textbf{X\#\#\#}). All these spectra are also summarised in Table \ref{tab:spectra}.

\begin{table*}
    \centering
    \caption{Luminosity by band and associated energy for each spectrum tested. Units are $\mathrm{ergs~s^{-1}}$. The spectra are all discrete with a single energy bin per band (listed in brackets) except for E, which is continuous - in these cases the listed luminosity is that integrated over the range indicated.}
    \begin{tabular}{c|c|c|c|c|c}
        \hline
         %Band    &   \citet{Ercolano_2009}   &   \citetalias{Wang_2017} \\
         Key            &   W   &   U   &   S\#\#\#   &  E  &   X\#\#\# \\
         Description    &   Fiducial   &   UV only   &  Soft X-ray   &   \citet{Ercolano_2009}  &   High $L_X$\\
         \hline
         Soft FUV       &   $5.04\times10^{31}$ & $5.04\times10^{31}$ & $5.04\times10^{31}$ & - & $5.04\times10^{31}$
         \\
         &  (7 eV) &  (7 eV) &  (7 eV) & &  (7 eV)\\

         Lyman-Werner   &   $3.07\times10^{29}$ &  $3.07\times10^{29}$ & $3.07\times10^{29}$ & $1.6\times10^{31}$ & $3.07\times10^{29}$
         \\
         &  (12 eV) &  (12 eV) &  (12 eV) & (11.27-13.6 eV) &  (12 eV)\\
         
         EUV            &   $2\times10^{31}$ & $2\times10^{31}$ & $2\times10^{31}$ & $2\times10^{31}$ & $2\times10^{31}$
         \\
         &  (25 eV) &  (25 eV) &  (25 eV) & (13.6-100 eV) &  (25 eV)\\
         
         X-ray          &   $2.56\times10^{30}$ & - & $2.56\times10^{30}$ & $1.6\times10^{31}$ & $1.6\times10^{31}$
         \\
         &  (1000 eV) & &  (\#\#\# eV) & (100-12000 eV) &  (\#\#\# eV)\\
            \hline
         %Total          &   $4\times10^{30}~\mathrm{erg~s^{-1}}$   & $5.5267\times10^{31}~\mathrm{erg~s^{-1}}$ \\
         %Ionising       &   &   $5.16\times10^{40}~\mathrm{s^{-1}}$
    \end{tabular}
    \label{tab:spectra}
\end{table*}

\subsubsection{Cooling}
We present simulations for two cooling models which we also summarise in Table \ref{tab:cooling}:
\begin{itemize}
    \item \textbf{Full}, which includes all the mechanisms inherent to \textsc{mocassin}. We also added atomic data for \ion{S}{I} and the first 30 energy levels \ion{Fe}{II} (enough to include the most accessible states) from CHIANTI \citep{Dere_1997,Dere_2019} so that collisionally excited forbidden line radiation from these species is included, in particular the transitions producing the [\ion{S}{I}] $25~\mu\mathrm{m}$ and [\ion{Fe}{II}] $26~\mu\mathrm{m}$ lines included by \citetalias{Wang_2017} (though they only found the former to be particularly significant).
    \item \textbf{IRFLNoH}, in which the only forbidden lines included are those longward of $10~\mu\mathrm{m}$ in order to eliminate cooling from optical and near infrared lines which were not modelled by \cite{Wang_2017} and in which the cooling by the Lyman alpha and beta lines due to collisionally excited hydrogen is also turned off in order to mimic the attenuation of the line core.
\end{itemize}
Note that we expect the Full cooling to be a better representation of the physical reality, while the more restricted IRFLNoH is an exercise designed to bring the \textsc{mocassin} treatment closer to that of \citetalias{Wang_2017}.
Nevertheless, both models still exclude some processes - in particular molecular cooling and hydrodynamical cooling - which are not currently implemented within \textsc{mocassin}. We discuss the potential impact of these in Section \ref{sec:cooling}.

\begin{table}
    \centering
    \caption{Summary of cooling processes included in our cooling models versus \citetalias{Wang_2017}.}
    \begin{tabular}{r|l|p{1.5cm}|p{3cm}}
         \hline
         Process        &   Full&   IRFLNoH &   \citetalias{Wang_2017} \\
         \hline
         Lyman Alpha    &   Yes &   No  &   Yes (escape probability)  \\
         Metal CELs     &   Yes &   Yes (>$10~\mu\mathrm{m}$ only)    & [\ion{C}{II}] $158~\mu\mathrm{m}$, [\ion{O}{I}] $63~\mu\mathrm{m}$, [\ion{S}{I}] $25~\mu\mathrm{m}$, [\ion{Si}{II}] $35~\mu\mathrm{m}$, [\ion{Fe}{I}] $24~\mu\mathrm{m}$, [\ion{Fe}{II}] $26~\mu\mathrm{m}$ \\
         Recombinations &   Yes &   Yes &   Yes \\
         Molecules      &   No  &   No  &   $\mathrm{H_2}$, $\mathrm{OH}/\mathrm{H_2O}$ and $\mathrm{CO}$ \newline rovibrational\\
         \hline
    \end{tabular}
    \label{tab:cooling}
\end{table}

\subsubsection{Elemental abundances}
For the elements considered by \citetalias{Wang_2017}, we used the same abundances\footnote{$\mathrm{He/H}=0.1$, $\mathrm{C/H}=1.4\times10^{-4}$, $\mathrm{O/H}=3.2\times10^{-4}$, $\mathrm{Si/H}=1.7\times10^{-6}$, $\mathrm{S/H}=2.8\times10^{-5}$, $\mathrm{Fe/H}=1.7\times10^{-7}$.}, with all other elements set to zero. For the elements that overlap with \citet{Ercolano_2009} (H, He, C, O, Si, S) these values are identical; in addition \citetalias{Wang_2017} include Fe but not N, Ne or Mg.
There can be reasonably significant, observable, emission from the omitted elements: we tested the difference including them would make to our results, ultimately finding it made qualitatively little difference to the overall conclusion.

Assuming an atomic/ionic composition, the mean molecular weight $\mu$ of the gas can be calculated as
\begin{equation}
    \mu = \frac{\sum_i m_i A_i}{\sum_i A_i + n_{\rm e}/n_{\rm H}}
    \label{eq:mu}
    ,
\end{equation}
where $m_i$ are the atomic masses relative to hydrogen, $A_i$ are the atomic abundances relative to hydrogen and $n_{\rm e}/n_{\rm H}$ is the ratio of the free electron density to the hydrogen density. For a neutral atomic gas of our adopted composition, $\mu=1.287$, though the value can be somewhat lower in regions of significant ionisation.

\subsubsection{Dust}
We assume a single grain population of $5\text{\r{A}}$ with the "Car\_90" composition from \textsc{MOCASSIN}'s library of dust datafiles which represents a neutral carbon grain in the form of a PAH/graphitic solid, the closest available to the PAH assumption of \citepalias{Wang_2017}.

\section{Example Temperature Structure}
\label{sec:fiducial}
\begin{figure*}
    \centering
    \includegraphics{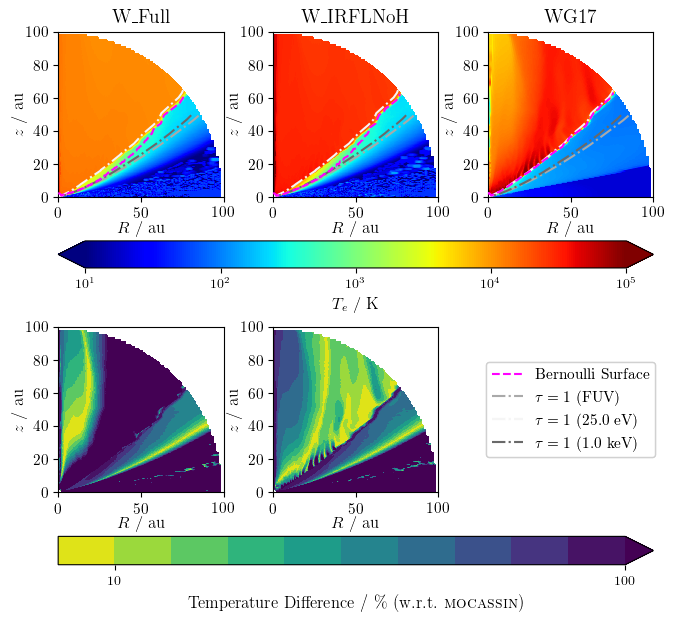}
    \caption{Comparison of the temperature structure obtained using \textsc{mocassin} with that of \citetalias{Wang_2017}. The left-most column shows the full cooling model, while the central column shows the restricted 'IRFLNoH' model in which cooling from Lyman lines and optical/NIR collisionally excited lines are turned off. The top row shows the temperatures, while the second row illustrates the percentage difference between the fiducial model of \citetalias{Wang_2017} and our simulations. In the bulk of the wind region the IRFLNoH cooling model has smaller temperature differences as indicated by the lighter colours. The pink dashed line indicates the surface where the Bernoulli function becomes positive while the dot-dashed lines (clockwise from the z axis) are the $\tau=1$ surfaces for EUV, X-ray and FUV respectively.}
    \label{fig:tempFiducial}
\end{figure*}

To provide context for our results, we first investigate whether our \textsc{mocassin} calculations can reproduce the temperature structure found by WG17 when we employ the same input spectrum (i.e. EUV, FUV and $1000~\mathrm{eV}$ X-rays) and how this is impacted by the cooling rates.
Therefore in Fig. \ref{fig:tempFiducial} we present the temperature structure in the simulations with spectrum W for the two different cooling models described in Section 2.2.2).
Based on the penetration of different radiation we can delineate the resulting temperature structure into three broad regions - the wind, the warm disc and the cold disc - which we discuss in turn before demonstrating how we use the temperatures to determine where a wind can be launched.

%\cathie{Comment (Addressed?): can the following paragraph be shortened to make the main point that the \it{locations} of the ionisation fronts is very similar in the three upper  panels of Fig. 3, indicating that, as you say, the Monte Carlo radiative transfer solution by \textsc{mocassin} is consistent with the ray tracing conducted by \citetalias{Wang_2017}?}

In the models of \citetalias{Wang_2017}, the wind and disc are divided by an ionisation front (IF) where a sharp density contrast is seen (Fig. \ref{fig:dens}).
The locations of our IFs agree well with \citetalias{Wang_2017}, indicating that the Monte Carlo radiative transfer solution by \textsc{mocassin} is consistent with the ray tracing conducted by \citetalias{Wang_2017} and suggestive of a self-consistent EUV-driven solution. 
This can be seen in terms of the penetration depth of EUV with the $\tau=1$ surface marked in each case with the light dash-dotted line: being relatively close in frequency to the first ionisation energies of hydrogen and helium, the EUV has the largest cross-section for photoionisation and hence the smallest penetration column of $N_{\ion{H}{I}} \sim 5\times10^{17}~\mathrm{cm^{-2}}$.
The low densities of the EUV-driven wind, and its high levels of ionisation, are key to allowing the EUV to reach this far, in contrast to the higher density X-ray driven winds into which the EUV does not penetrate very far \citep{Owen_2012}.

%Nevertheless, the low densities in the EUV-driven wind of \citetalias{Wang_2017} mean that the EUV in both simulations also penetrates (as indicated by the lightest dot-dashed line) to the location of the ionisation front where a sharp density contrast is seen. This is in 
%We first turn our attention to the wind region that all the radiation must pass through.

The wind region penetrated by the EUV is hot and approximately isothermal: for the Full cooling model at around $10^4~\mathrm{K}$ and for the IRFLNoH cooling model at around $3\times10^4~\mathrm{K}$.
The latter model yields temperatures that are close (within around 30\%) to those obtained by \citetalias{Wang_2017}: this  is  as expected given that the model omits cooling processes also omitted by \citetalias{Wang_2017}\footnote{We also tried to achieve agreement with the hot temperatures of \citetalias{Wang_2017} by more measured means such as removing only a) the Lyman lines, b) certain metals or c) even individual lines from the cooling, but only the IRFLNoH model combining several of these was able to produce the $\sim 3\times10^4~\mathrm{K}$ wind temperatures.}.  On the other hand 
%In the former case, the temperature is limited by the onset of cooling via Lyman radiation from any remaining neutral hydrogen - for which the cooling rate is a strong function of temperature and acts as an effective thermostat. 
%In making a comparison to the temperatures derived by \citetalias{Wang_2017}, we can identify two subregions: in the bulk of the wind at $R\gtrsim 20~\mathrm{au}$, the hotter temperatures of the IRFLNoH are a better fit, agreeing to within around 30\%.
%This implies that the inefficiency of the cooling processes that are excited at several thousands of kelvin are responsible for higher temperatures of this model in this region.
%However, 
within $R\lesssim20~\mathrm{au}$ and at $z\gtrsim20~\mathrm{au}$, we see that \citetalias{Wang_2017} find a cooler lobe of temperatures much closer to the $10^4~\mathrm{K}$ of the Full Model. %We explore the origin of these differences in Section \ref{sec:cooling} but broadly speaking the unusually hot wind temperatures of \citetalias{Wang_2017} can be explained by their choice of atomic cooling processes\footnote{We also tried to achieve agreement with the hot temperatures of \citetalias{Wang_2017} by more measured means such as removing only a) the Lyman lines, b) certain metals or c) even individual lines from the cooling, but only the IRFLNoH model combining several of these was able to produce the $\sim 3\times10^4~\mathrm{K}$ wind temperatures.}
We suggest in Section \ref{sec:cooling} that the origin of these cooler temperatures is adiabatic cooling which is neglected in our (radiative equilibrium) models.
%and our IRFLNoH model must be missing some cooling in the cooler lobe which must have similar magnitude, but different origin, to the additional terms in the Full model.

While in principle hotter wind temperatures act to make the wind more highly ionised and therefore more transparent to radiation, we find little change in the location of the ionisation front regardless of cooling and wind temperature.
This is because the wind is primarily photoionised rather than thermally ionised so the hotter temperatures make little difference to the transparency of the wind. Moreover the penetration depths of the X-ray and FUV frequencies are sufficiently large that any modification to the neutral column in the EUV-heated wind region has negligible effect.

Beyond the ionisation front, both of our simulations have broadly the same appearance. First we come to a warm $\sim 1000~\mathrm{K}$ region heated by both FUV and X-ray. We see that this region is generally warmer than it was found to be by \citetalias{Wang_2017} by a few $100~\mathrm{K}$;
%\cathie{Comment: could this be to do with molecules?}
this implies our models either have some additional heating or are missing some cooling - we suggest in Section \ref{sec:molecules} that this is likely the result of molecular cooling. There is little difference between the cooling models as these temperatures are not generally warm enough to significantly excite the optical lines that are turned off in the IRFLNoH model.

The FUV has an opacity largely dominated by dust absorption\footnote{Though there are also significant contributions from carbon and sulfur photoionisation.}, and for the grains in question, can penetrate a column of roughly $2\times10^{22}~\mathrm{cm^{-2}}$, while the high energy $1~\mathrm{keV}$ X-rays can reach around $10^{22}~\mathrm{cm^{-2}}$. Therefore, the FUV and X-ray reach similar depths in this case.
Beyond this point the temperatures appear to decline and tend towards better agreement with \citetalias{Wang_2017}.
The midplane of the disc is dark to all the bands of radiation included. The \textsc{mocassin} temperatures in this region are noisy due to low photon counts and the low temperature behaviour of the algorithm applied to find equilibrium.

In summary then, our  IRFLNoH model, which turns off a number
of atomic cooling channels, reasonably reproduces the temperature
structure found by \citetalias{Wang_2017}, albeit with slightly
warmer conditions below the ionisation front.

For a given temperature profile, a wind can be launched wherever the material is hot enough to be unbound and overcome the gravitational potential $\Phi$.
%Assuming an adiabatic equation of state, \citetalias{Wang_2017} show
This can be described in terms of the volume in which the Bernoulli parameter is positive \citep{Liffman_2003,Wang_2017}
\begin{equation}
    B := \frac{v^2}{2} + \frac{\gamma}{\gamma-1} \frac{P}{\rho} + \Phi > 0
    .
\end{equation}

Since we do not solve the hydrodynamics, we cannot determine the velocities $v$ for our simulations. However, we can assume that they are initially subsonic \citep{Clarke_2016,Sellek_2021} at the base of the wind and so the poloidal kinetic energy term $v_p^2/2$ is small compared to the thermal term $P/\rho$, while the azimuthal kinetic energy term $v_\phi^2/2 \sim \frac{\Phi}{2}$. Thus, we define the Bernoulli surface as the location where
\begin{equation}
    T = T_{\rm Bern} := \frac{1}{5} \frac{GM_* \mu m_H}{k_B r} \approx 27600~\mathrm{K} \left(\frac{r}{\mathrm{au}}\right)^{-1} \frac{\mu}{1.287}
    \label{eq:TBern}
    ,
\end{equation}
assuming that the wind is mainly atomic and so $\gamma=5/3$ and where the typical value is given assuming a star of $1~M_{\sun}$. When assessing whether a given cell has a temperature above $T_{\rm Bern}$, its local value of the mean molecular weight $\mu$ - calculated from the local ionisation fraction as per equation (\ref{eq:mu}) - is used.
Up to a numerical factor, this is similar to the escape temperature formalism of \citet{Owen_2012} \citep[though see][for a discussion about how this does not completely accurately capture the temperature at the wind base]{Picogna_2021}.

The Bernoulli surface is plotted as the magenta dashed line on Fig. \ref{fig:tempFiducial}. In the case of \citetalias{Wang_2017}'s temperature profile it lies along the ionisation front, coincident with the $\tau=1$ surface of EUV. This is strongly indicative of an EUV-driven wind. In our simulations, for $R \gtrsim 50~\mathrm{au}$, we also see a good agreement between these two, though the temperature gradient at this location is less sharp so there is a small difference.
Within $10-50~\mathrm{au}$, the Bernoulli surface dips down below the ionisation front, implying that in our simulations, the combination of FUV and X-ray is capable of heating the material to a hot enough temperature to drive a wind (though only at relatively low columns and not all the way down to their $\tau=1$ surfaces).
%\cathie{is the Bernoulli surface significantly lower in the IRFLNOH case? This isn't obvious from the Fig.....)}.
In the IRFLNoH case, this dip extends to slightly smaller radii due to the reduced cooling around the higher inner-disc $T_{\rm Bern}$. Viewed another way, the innermost limit is on the order of the gravitational radius $r_G = \frac{GM_*}{c_S^2}$, which is smaller for the hotter wind temperatures. 

Our \textsc{mocassin} simulations with input spectrum matching that of \citetalias{Wang_2017} thus produce thermal structures that
corroborate the conclusion of \citetalias{Wang_2017} that $1000~\mathrm{eV}$ X-rays are not able to drive a wind from the outer disc regardless of any differences in cooling processes. In the inner disc, the simulations indicate a possible minor role for FUV/X-ray wind launching but the Bernoulli surface is only modestly below the ionisation front. We are now in a position to assess how this conclusion is affected when we vary the input X-ray spectrum employed.
%Thus we conclude that for the cooling included in \textsc{mocassin}, there are some radii where an FUV/X-ray driven wind might be possible. However, at large radii, $1000~\mathrm{eV}$ X-rays are not able to drive a wind even without the contributions of any cooling not included in \textsc{mocassin}.

\section{Role of X-ray frequency}
\label{sec:varybands}

\begin{figure*}
    \centering
    \includegraphics{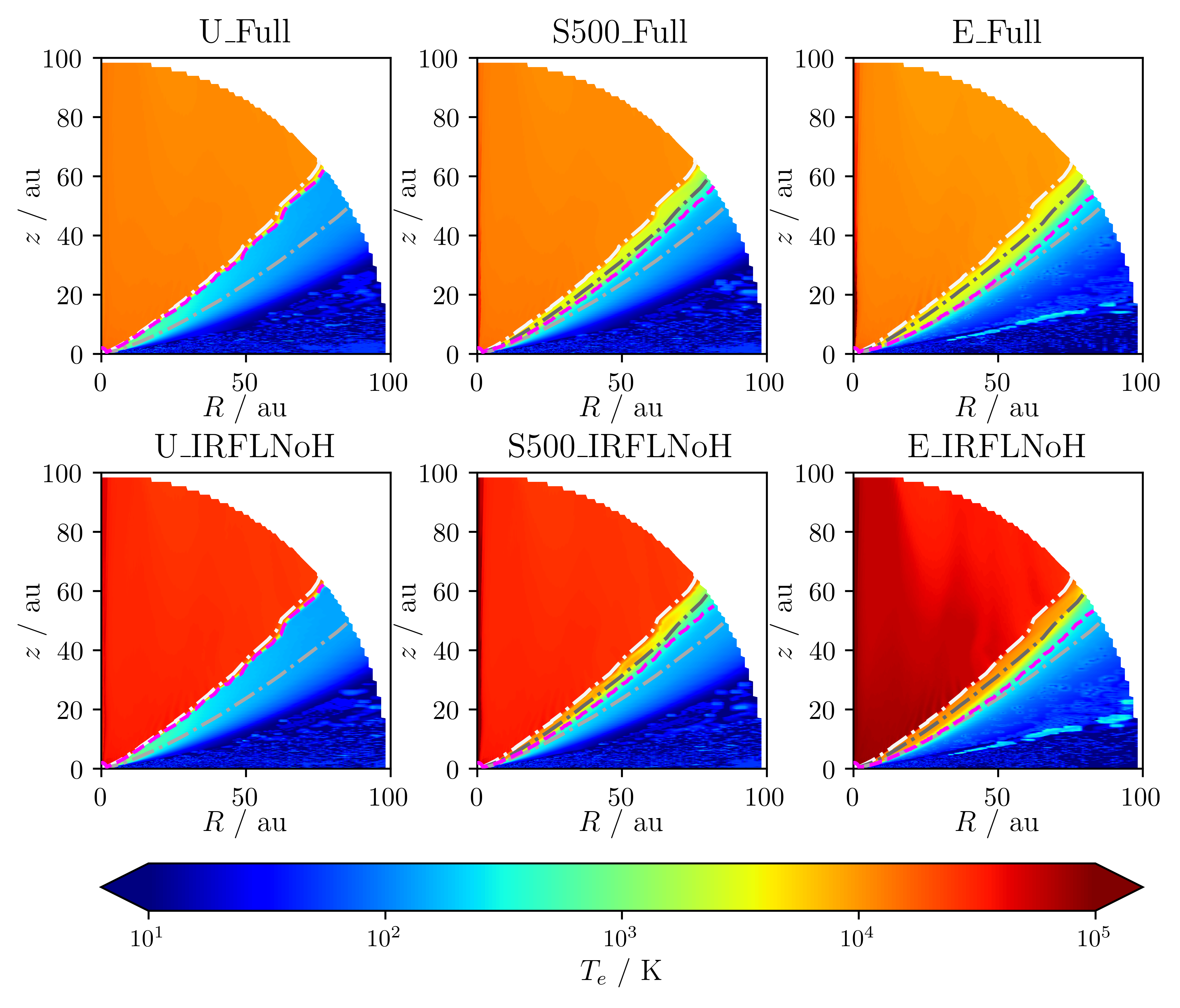}
    \caption{Comparison of the temperature structure obtained for different combinations of spectra and cooling model. From left to right: the UV-only spectrum, a spectrum with $500~\mathrm{eV}$ X-ray and the spectrum of \citet{Ercolano_2009}. In each case the pink dashed line indicates the surface where the Bernoulli function becomes positive while the dot-dashed lines represent $\tau=1$ surfaces for EUV, $500~\mathrm{eV}$ X-ray and FUV.}
    \label{fig:tempSpec}
\end{figure*}

Having understood how the chosen cooling models give rise to the temperature profiles, we wish to further elucidate the role that the X-rays are playing in our simulations, and examine the potential of different X-ray energies to drive a wind.
In particular we will seek the most effective radiation and so we are interested in whether there are X-ray bands that can heat a larger column than the EUV. 

Therefore, as a control, we first run a pair of simulations with the X-rays removed that is UV-only (simulations U\_Full and U\_IRFLNoH) and present their temperature profiles in the first column of Fig. \ref{fig:tempSpec}.
Removal of X-rays  makes fairly little difference to the overall picture of a $10^4~\mathrm{K}$ ($3\times10^4~\mathrm{K}$) wind for the Full (IRFLNoH) cooling model.
On the other hand, we see cooler temperatures below the ionisation front, with the remaining heating provided mainly by FUV photoionisation of carbon. The difference to the fiducial simulations confirms the role of X-ray in heating this region in those earlier models through photoionisation of hydrogen and helium.
%\cathie{Comment: the next sentence is true but maybe confusing to the reader since it comes in a discussion of models without X-rays. Surely the place to discuss missing coolants below the IF in our simulations is the previous section when we talk about the slightly higher temperatures that we get below the IF cf W17?}
Nevertheless while closer to those found by \citetalias{Wang_2017}, the temperatures are still a little too hot, strengthening the argument for missing coolants in that region (as opposed to say, uncertainties in X-ray heating efficiency).
%even without X-rays these cooler temperatures are actually closer to , implying that an additional source of cooling may be needed here to effectively offset the X-ray heating contribution.
These temperatures are, however, sufficiently low that the Bernoulli surface no longer dips down but follows the ionisation front at all radii. Therefore at these luminosities, FUV alone is not able to drive a wind from below the ionisation front. That said, a higher FUV luminosity \citep[e.g.][]{Nakatani_2018a,Nakatani_2018b}, different FUV spectrum, different assumptions about the nature of the dust, or inclusion of molecular heating (e.g. FUV pumping of H\textsubscript{2}) may allow for more significant FUV heating (e.g. through the dust photoelectric effect), potentially  sufficient to launch a wind. Exploring the role of FUV further is beyond the scope of this work.

Now we proceed to vary the frequency of the radiation band from $100~\mathrm{eV}$ to $900~\mathrm{eV}$ in steps of $100~\mathrm{eV}$. The aim of this exercise is to determine the impact of using a single band - and correspondingly the choice of its energy - on X-ray driving of winds. We thus keep the luminosity constant while doing so.

In the second column of Fig. \ref{fig:tempSpec} we depict the temperature structure for the runs with $500~\mathrm{eV}$ which present the largest contrast with the 1 keV results discussed hitherto.
%These are fully explored below in \ref{sec:effective}; here for maximum contrast we show the results for $500~\mathrm{eV}$ in the second column of \ref{fig:tempSpec}. 
For both of our cooling models, X-rays of this energy are clearly able to heat material comfortably below the ionisation front to escape and thus drive a wind from all radii. It is therefore clear that the choice of X-ray frequency is a key parameter affecting whether X-rays can heat material beyond the EUV ionisation front sufficiently to drive a wind. %The column which the X-rays can heat to escape as a function of energy is considered on more depth in Section \ref{sec:toymodel}.

\subsection{Which radiation is most effective?}
\label{sec:effective}
\begin{figure*}
    \centering
    \includegraphics{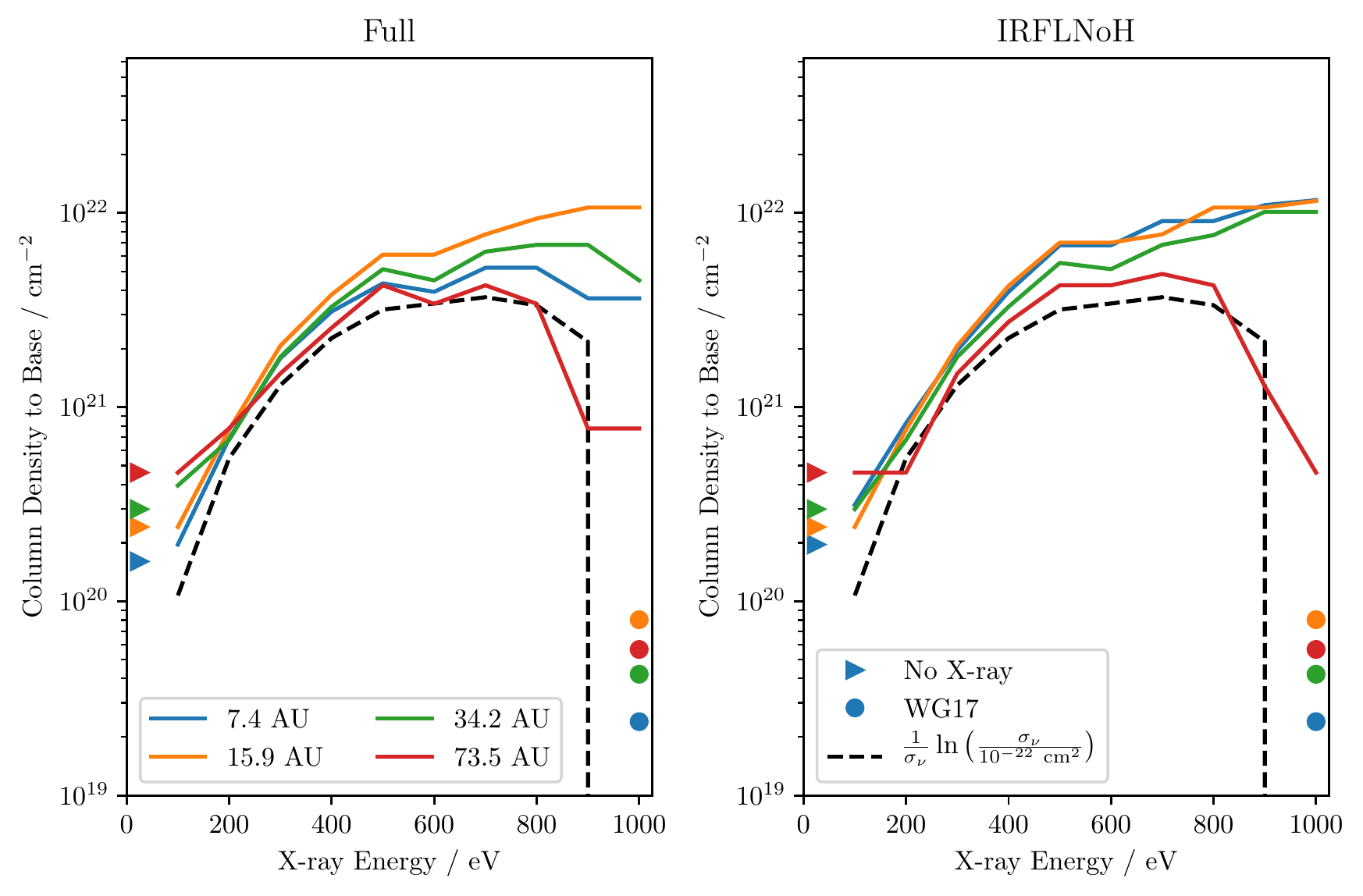}
    \caption{The \ion{H}{I} column to the Bernoulli surface for each energy of X-ray at selected radii (solid lines). Simulations with the Full cooling model are shown on the left hand panel and the IRFLNoH cooling on right. In each case, the triangles and circles represent the corresponding values for the UV-only spectrum and \citetalias{Wang_2017}'s temperature field respectively. The dashed line is equation (\ref{eq:Neff}) for $\epsilon_c=10^{-22}~\mathrm{cm^{2}}$.}
    \label{fig:columnE}
\end{figure*}

To illustrate which X-ray bands can effectively heat a larger column than the EUV, we plot in Fig. \ref{fig:columnE} the neutral hydrogen column density to the Bernoulli surface achieved by each energy for our S\#\#\# simulations. We show this dependence for 4 different radii as the solid lines. In addition, the triangles of corresponding colour (plotted at $25~\mathrm{eV}$) mark the column at the Bernoulli surface at the same radii in our UV-only simulations and the circles (plotted at $1000~\mathrm{eV}$) likewise for \citetalias{Wang_2017}'s fiducial model.

Above around $800~\mathrm{eV}$, we see that at large radii, the X-rays cannot heat a greater column than the EUV. Moreover, for all radii, at the lowest energies, the column heated by the X-ray is not much greater than the EUV as these frequencies are quite strongly absorbed. The most effective choices for a single X-ray energy that will heat the largest column are those in the range $500-700~\mathrm{eV}$, depending slightly on the radius in question.
We explore the shape of these curves with a simple model in Section \ref{sec:toymodel}, illustrated here with the black dashed line (equation \ref{eq:Neff} for a fiducial value of $\epsilon_c=10^{-22}~\mathrm{cm^2}$).
Note that since we control $L_X$, although in each case our spectrum is a effectively a delta function, Fig. \ref{fig:columnE} also in effect gives the relative contribution in a flat spectrum where each band has luminosity of $2.56\times10^{30}~\mathrm{erg~s^{-1}}$. For now we will proceed to discuss these as indiviudal choices (i.e. as delta function spectra attempting to capture the whole spectrum), but will return in Section \ref{sec:full_spectrum} to examine the effect of a realistic spectral shape in determining what is genuinely representative.
%The shape of the curves from the simulation data are also broadly in agreement with our toy model.

The choice of cooling model makes fairly little qualitative difference to these results; in part because the cooling rates between them are not so different for typical values of $T_{\rm Bern}$.
The biggest difference between the two panels of Fig. \ref{fig:columnE} is seen for $7~\mathrm{au}$ where $T_{\rm Bern}$ is high enough for the omission or inclusion of atomic cooling channels to affect the temperature attained.
In either case, an X-ray driven wind can be launched here and given the modest contribution to the total wind mass loss rate from such small radii, the correct treatment of optical forbidden line cooling and Ly $\alpha$ (and $\beta$) cooling is not an important factor in determining X-ray driven mass loss.
%where no X-rays are able to launch a wind with full cooling, but all frequencies are somewhat potent for model IRFLNoH. This has to result from the diverging in the cooling rates at higher temperatures, showing that the choice of cooling processes can have an important impact on the location of the base.

Instead we conclude that the limited role for X-rays relative to UV in the simulations of \citetalias{Wang_2017}, predominantly reflects
%\cathie{does not depend on differences in cooling processes included in their work but instead reflect}
the fact that $1000~\mathrm{eV}$ X-rays are too hard - and so interact too weakly with the disc gas - to heat it sufficiently to drive a wind on their own, regardless of the differences in cooling processes.
%\cathie{Just checking I've understood this: the reason that WG17 is cooler below the IF than U is surely because of molecular cooling. So WG17 has a particularly small column because of the combination of having only 1000 keV X-rays and molecular cooling?}
That said, in absolute terms, the columns at the base in the simulations by \citetalias{Wang_2017} are only $2-8\times10^{19}~\mathrm{cm^{-2}}$. Whereas, despite the Bernoulli surface in our UV-only simulations lying very close to that of \citetalias{Wang_2017}, the column at an equivalent radius can be up to $\sim10$ times higher at $1-5\times10^{20}~\mathrm{cm^{-2}}$. The origin of this behaviour is that the temperature gradient is shallower in our UV-only simulations because they are hotter below the base than found by \citetalias{Wang_2017}. Thus, the Bernoulli temperature is reached at a slightly lower height, below the IF. However, since the base is only mildly flared, photons reach it at a very glancing angle - the distance travelled below the \citetalias{Wang_2017} IF at $n_{\rm \ion{H}{I}}\sim10^{6}~\mathrm{cm^{-3}}$ is thus considerable.

We cannot directly measure mass loss rates from our models as we have not performed hydrodynamic simulations to adapt the density and velocity fields to be consistent with our different temperatures.
However since the amount of mass loss determines how much material the radiation has to pass through to reach the wind base, it is reasonable to assume that $\dot{M}\propto N$ (i.e. we are assuming that $N \propto n_{\rm base}$ and $\dot{M} \propto n_{\rm base}$).
Therefore, we would expect the higher columns in the UV-only simulations to translate into a similar factor $\sim10$ boost in the mass loss rates.
We ascribe this difference to additional cooling in the model of \citetalias{Wang_2017}; indeed, when they produced a setup closer to \citet{Owen_2010} (their OECA analog) by turning off some of this cooling, they did see a mass loss rate that was higher by a factor of $4-5$. 
Note that since in the outer disc, which typically dominates the mass loss rates, X-rays cannot heat a larger column than EUV, we would expect the mass loss rates of a $1000~\mathrm{eV}$ X-ray simulation to be only marginally higher than an UV-only one, as observed by \citetalias{Wang_2017}.

In this context, we estimate that if one wishes to use a single energy to represent the X-rays, moving to $\sim500~\mathrm{ev}$ would increase the mass loss rate by a factor $\sim4-6$ over that found in a simulation driven by UV-only. As we will discuss further in Section \ref{sec:full_spectrum}, the shape of the spectrum controls whether such energies are present in sufficient numbers to be representative.

%cathie{Re the following sentence: while it's not to be doubted that turning off some cooling might increase the mass loss

%By contrast, our UV-only simulations, which are the closest in temperature to \citetalias{Wang_2017} but still somewhat hotter, reach just under 10 times higher.
%Moreover our comparable X-ray simulations reach $0.5-5\times10^{21}~\mathrm{cm^{-2}}$, 1-2 orders of magnitude larger depending on radius.
%ates, it doesn't really back up the idea that the heated column scales linearly with the mass loss rate since this wasn't evaluated
%y WG17....}.
%n increase in mass loss rate was similarly seen in the experiment performed by \citetalias{Wang_2017} in producing a setup closer to \citet{Owen_2010}: their OECA10 analog where they turn off some of their additional cooling had a significantly higher mass-loss rate of $11.2\pm4.2\times10^{-9}~M_\odot\mathrm{yr^{-1}}$ compared to $2.5\pm0.2\times10^{-9}~M_\odot\mathrm{yr^{-1}}$ for their fiducial model. \cathie{Next sentence: doesn't this just repeat what's written above?}.Moreover, moving to a more efficient X-ray energy could provide a further boost to the mass loss rates; the most efficient ones can consistently heat $2-5\times10^{21}~\mathrm{cm^{-2}}$, about an order of magnitude higher than if there were no X-ray at all.

\subsection{Explanatory Model}
\label{sec:toymodel}
As discussed before, we assume that a wind is launched when gas is heated above $T_{\rm Bern}$ (equation \ref{eq:TBern}). We now consider a toy model for whether monochromatic X-ray radiation of frequency $\nu$ can launch a wind on its own.

Assuming that all X-ray radiation absorbed goes into heating the gas, the heating rate per unit volume can be written in terms of the geometrically diluted and attenuated X-ray flux, $F_X$ ($ = \frac{L_X}{4\pi r^2} e^{-N\sigma_\nu}$), local gas density, $n$ and photoionisation cross section, $\sigma_\nu$, as $F_X n \sigma_\nu$. We assume that $\sigma_\nu$ is independent of temperature since the gas is predominantly photoionised rather than thermally ionised\footnote{In practice, the cross section will depend somewhat on temperature, since the ionisation state of the absorbing material depends on the recombination coefficient $\alpha$ which is temperature dependent.}.

However it is important to note that this form for the heating is an overestimate.
Firstly, it neglects the fact that some of the energy is used up in overcoming the ionisation energy of the electrons; for X-ray ionisation of hydrogen (and to a lesser extent helium) this is only a small correction $\lesssim 10$ per cent, but could be more significant for metals such as oxygen, where inner shells have ionisation energies in the 100s of eV.
Moreover, it neglects further losses due to secondary ionisation by the Auger effect in the heavier elements, and similarly the possibility of that the energy carried by the photoelectron may be lost to further ionisations or collisional excitation before it can thermalise \citep[e.g.][]{Maloney_1996}. 
On the other hand, \textsc{mocassin} treats the X-ray heating more self-consistently, accounting for losses to secondary ionisation and excitation as a function of ionisation fraction using the fits of \citet{Shull_1985}; for high levels of ionisation, the chance of thermalising through electron-electron collisions becomes greater than the chance of ionising or exciting neutral hydrogen, so the heating fraction $f_X\to1$. However in gas with a largely atomic composition, it can be on the order of 10 per cent. Hence we will scale the heating term in our thermal balance by $f_X$.
%, which results in somewhat higher efficiencies, even in largely neutral material.

The predominant cooling effects are two-body processes, requiring a collisional excitation between an electron or neutral and an ion or neutral - assuming these are below their critical densities (above which collisional de-excitation dominates over radiative de-excitation), the cooling rate per unit volume for each can be written as $n^2 \Lambda_{n,i}(T)$ (where the subscript $n$ here indicates that $\Lambda$ is per particle per number density), and hence the total cooling rate is of the same form $n^2 \Lambda_n(T)$ ($\Lambda_n(T) = \sum_i \Lambda_{n,i}(T)$).

Assuming thermal equilibrium, we may set the heating and cooling to equal, and rearrange to find
\begin{equation}
    \Lambda_n(T) = \frac{f_X}{4\pi} \frac{L_X}{n r^2} \sigma_\nu e^{-N \sigma_\nu} = \frac{f_X\xi}{4\pi} \sigma_\nu e^{-N \sigma_\nu}
    \label{eq:toymodel}
    ,
\end{equation}
where we can identify the ionisation parameter $\xi = \frac{L_X}{n r^2}$ \citep{Tarter_1969,Owen_2010}.
We thus see that the relationship between the ionisation parameter and temperature depends on the column density at the material $N$ \citep{Picogna_2019} as well as the frequency of radiation involved \citep[c.f. the spectral shape][]{Ercolano_2021}.

A simple consequence of equation \ref{eq:toymodel} is that assuming $\Lambda(T)$ is a monotonically increasing function of $T$, the highest temperatures at a given column are produced by that radiation, across all frequencies, for which $\tau=N\sigma_\nu=1$. More energetic radiation is more deeply penetrating so is simply not absorbed well enough locally to deposit much energy into the material. Conversely, less energetic radiation is more easily absorbed and so has been too strongly attenuated by the time is reaches the column $N$.
One can thus replace the frequency dependent terms with an ``efficiency" (with dimensions of a cross section) for a given frequency of radiation: 
\begin{equation}
    \sigma_\nu e^{-N \sigma_\nu} \to \epsilon_\nu
    \label{eq:efficiency}
    .
\end{equation}
Note that while we here call this the efficiency, other works refer to $f_X$ using the same name; in practice both determine the ability of the X-rays to heat the gas and effects such as overcoming the ionisation energy will mean that the real efficiency $< \sigma_\nu e^{-N \sigma_\nu}$. The constraints on cross sections capable of heating that we now proceed to derive are only strengthened by these effects.

The requirement that $T \geq T_{\rm Bern}$ in the wind gives us a requirement on the minimum efficiency of the X-rays\footnote{Note that since we assumed the cooling depends quadratically on density, the local density explicitly appears in this formula through $\xi$; however if the density is sufficiently high for a linear dependence, our approach still works with a different definition of the cooling rate, and $\epsilon_c$ will become independent of $n$.}
\begin{equation}
    \epsilon_\nu \geq \epsilon_c := \frac{4\pi \Lambda_n(T_{\rm Bern})}{f_X\xi}
    \label{eq:efficiencyc}
    .
\end{equation}

Given the X-ray luminosity $L_X$, stellar mass $M_*$, radius $r$,  local density $n$ and information about the ionisation states of each element (which controls $\Lambda(T)$ and $f_X$), we can determine a value for $\epsilon_c$ at each location in the density field. Note that the X-ray luminosity, cooling rate and heating fraction are degenerate in their effects on $\epsilon_c$, and so each may be changed to similar effect. We discuss them each in more depth in sections \ref{sec:highL}, \ref{sec:molecules} and \ref{sec:f_X} respectively.

From the definition of the efficiency, we can solve for the maximum column that radiation of a single frequency can heat to the required temperature
\begin{equation}
    N_{\rm max} = \frac{1}{\sigma_\nu}\ln\left(\frac{\sigma_\nu}{\epsilon_c}\right)
    \label{eq:Neff}
    ,
\end{equation}
where we expect that unbound material should exist anywhere that $N<N_{\rm max}$.

A necessary but not sufficient condition is therefore that $N_{\rm max}>0$, in which case our single choice of frequency must have $\sigma_\nu>\epsilon_c$ since otherwise even completely unattenuated radiation could not heat the wind. This imposes an upper bound on the X-ray energies that can heat the gas to the escape temperature. The highest column (at fixed $\epsilon_c$) is heated by radiation with $\sigma_\nu = e \epsilon_c$, such that $N = 1/\sigma_\nu = 1/(e\epsilon_c)$. At larger still values of $\sigma_\nu$, the column heated is moderately larger than $1/\sigma_\nu$ ($\tau\gtrsim1$), but is nevertheless a decreasing function of $\sigma_\nu$.

Note that correspondingly, the optical depth at the base is $\tau=1$ for the most efficient radiation. Since they heat inefficiently, higher energies are likely to be optically thin at the base, while the lower energies will be somewhat optically thick.
Therefore for radiation effective enough to drive an X-ray wind, we expect order unity optical depth at the base. Broadly speaking this means that the temperatures around the base are not declining purely due to increasing cooling from denser material but also by a decrease in heating as the radiation is attenuated too. The decrease in heating is the more important effect once $\tau > \frac{N}{nd}$ ($\sim 0.4-0.5$ below the base) where $d = n/\frac{\partial n}{\partial r}$. This means that an optically thin prescription using a single $\xi-T$ relation and assuming lower temperatures result only from lower densities \citep[e.g.][]{Owen_2012} will generally be less accurate than an attempt to account for column density or attenuation of radiation.

\subsubsection{Application to Results}
We first examine which of the X-ray radiation bands can heat a larger column than UV alone, by evaluating $\epsilon_c$ equation (\ref{eq:efficiencyc}) along the Bernoulli surface (i.e. the ionisation front) in the ultra-violet only model; to allow for comparison to \citetalias{Wang_2017}, we use $L_X=2.56\times10^{30}~\mathrm{erg~s^{-1}}$, and since EUV-heated gas is nearly completely ionised, we assume $f_X=1$.
We have argued that a necessary condition for effective X-ray heating below the ionisation front is that $\sigma_\nu > \epsilon_c$, so if the values of $\epsilon_c$ correspond to cross-sections in the X-ray regime then we expect some X-rays to be potent sources of heating at - and therefore somewhat below - the UV-only wind base and hence an X-ray driven wind to be possible.
%Since $\sigma_\nu > \epsilon_\nu$ (equation 4) and we require $ \epsilon_\nu \geq \epsilon_c$ then a necessary condition for .
%required using the results of the UV-only simulations; 
%\cathie{In order to establish the maximum column, $N_{max}$, to which X-rays of a given frequency can heat gas to $T_{Bern}$,
%  for each frequency, we then evaluate 
%$N_{max}$ using equation (6).}
%Since the aim is to understand which of the radiation bands can heat a larger column than UV alone, we evaluate the minimum efficiency $\epsilon_c$ required using the results of the UV-only simulations; 
%If these values correspond to c
%We therefore calculate $\epsilon_c$ along the Bernoulli surfaces shown in the first column of figure \ref{fig:tempSpec} and show the results in the left-hand panel of 

\begin{figure}
    \centering
    \includegraphics[width=\linewidth]{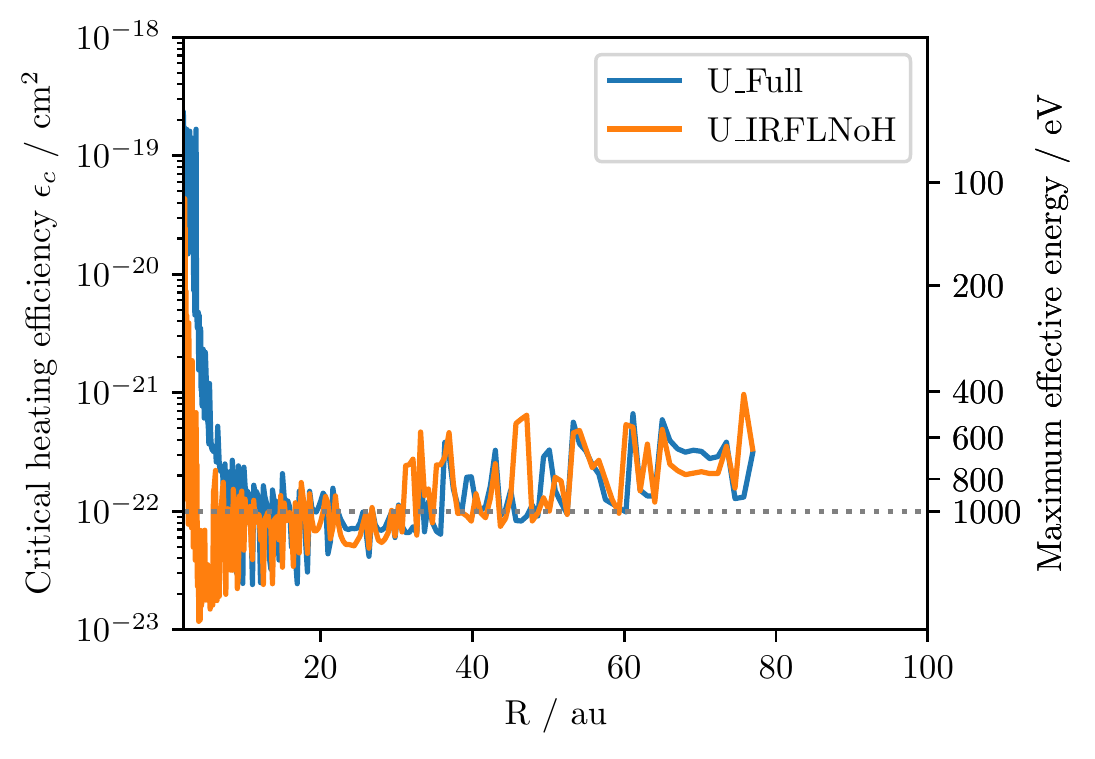}
    \caption{The critical efficiency required to overcome the local cooling along the Bernoulli surface in the U\_Full (blue) and U\_IRFLNoH (orange) simulations as a function of radius assuming $L_X=2.56\times10^{30}~\mathrm{erg~s^{-1}}$ \citepalias{Wang_2017} and $f_X=1$. The right-hand axis calibrates this scale in terms of the X-ray energy with cross-section equal to this value - any higher energy will have too low a cross-section to achieve the required efficiency. The maximum energy that is effective on its own at this luminosity is similar between the cooling models except for the inner $20~\mathrm{au}$. $1000~\mathrm{eV}$ as used by \citetalias{Wang_2017} is marked with the dotted line for reference.}
    %The right-hand panel includes for comparison, the simulations with spectrum W containing $1000~\mathrm{eV}$ - the efficiencies are only significantly affected below $\epsilon_c=10^{-22}~\mathrm{cm^2}$ corresponding to the cross-section of these X-rays (marked with dotted line).}
    \label{fig:efficiency}
\end{figure}

Thus, Fig. \ref{fig:efficiency} depicts the run of $\epsilon_c$
with radius along the Bernoulli surface in the UV-only model (Fig. \ref{fig:tempSpec}) while the right hand axis equates values of
$\epsilon_c$ with the value of the X-ray photon energy for which 
$\sigma_\nu= \epsilon_c$, this being the maximum energy for which
heating to $T_{\rm Bern}$ would be possible even in the case of no attenuation.

Except in the innermost parts of the disc, the values for $\epsilon_c$ derived are in the range $10^{-22}-10^{-21}~\mathrm{cm^{2}}$ and reach a minimum around $10-20~\mathrm{au}$. These correspond to the photoionisation cross sections of photons in the range $400-1000~\mathrm{eV}$\footnote{To perform this calculation we assume a neutral gas and use the opacities of \citet{Verner_1995,Verner_1996}; in practice the cross sections for photoionisation are only weakly affected by ionisation except in terms of the exact location of the ionisation energy.}. The shallow increase to larger radii is due to the effects of geometric dilution weakening the irradiating flux, though this is largely offset by the material being less tightly bound (with a lower $T_{\rm Bern}$ at which the cooling rates to be overcome are lower) and less dense.

We would thus expect that over much of the disc outside $\gtrsim40~\mathrm{au}$, $1000~\mathrm{eV}$ acting alone should not be able to launch an X-ray driven wind, though it is marginally able to do so around $\sim 20~\mathrm{au}$. Indeed, this is what was discussed in Section \ref{sec:fiducial} and illustrated in Fig. \ref{fig:columnE}.
It is likely that in practice the ability of these harder X-rays to launch a wind was assisted by the presence of FUV which has a similar cross section for absorption and thus is contributing to the heating at these columns.
%\cathie{I am wondering if the right hand panel doesn't confuse more than it reveals? The left hand panel takes a situation with no X-rays and then asks which are the X-ray energies that could heat more than this and at what radii. This seems very clear: you are assessing a range of different energies against the UV only model. The right hand panel is more subtle. You are not now comparing each X-ray band's ability to exceed the heating provided by 1000 eV X-rays. You already know from the left hand panel that 1000 ev X-rays canheat to the UV surface at r< 20 AU but not beyond this. Therefore you already know that the Bernoulli surface won't change between the 2 panels at r > 20 AU and therefore the $\epsilon_c$ values shouldn't change there. And you expect the Bernoulli surface to change a bit at r < 20 AU and so slightly change the range of energies that could heat to the Bernoulli surface. But it strikes me that whereas this plot is quite consistent with what we know from the temperature plots, it doesn't add any new diagnostic information, whereas the left hand panel does....?}
%In the right-hand panel of \ref{fig:efficiency} we perform the same calculation at the Bernoulli surface for the fiducial models. The efficiency is changed only in the region where a wind launched by $1000~\mathrm{eV}$ X-rays was seen to occur: the marked change in the efficiency around a value of $10^{-22}~\mathrm{cm^{2}}$ is indicative of a radius beyond where these X-rays "drop out" and are no longer relevant to wind launching.

On the other hand, energies $\lesssim 600~\mathrm{eV}$ should be able to launch an X-ray driven wind from the entire disc as was the case for the $500~\mathrm{eV}$ example shown in Fig. \ref{fig:tempSpec}. The mild increase of $\epsilon_c$ with radius implies that the maximum energy for effective X-ray heating should decrease mildly with increasing radius.
Assuming a best-case scenario that a low efficiency of $\epsilon_c\sim10^{-22}~\mathrm{cm^2}$ applies, we should expect that the deepest penetration is $N \sim \frac{1}{e\times10^{-22}~\mathrm{cm^2}} \sim 4\times 10^{21}~\mathrm{cm^{-2}}$ and is achieved for an energy of $\sim700~\mathrm{eV}$. As we move towards the outer disc and $\epsilon_c$ increases, we should see this peak decrease and shift to lower energies.

Based on these results we adopt a fiducial value of $\epsilon_c=10^{-22}~\mathrm{cm^2}$ and can then use equation (\ref{eq:Neff}) to calculate the maximum penetration depth as a function of energy. This is shown as the black dashed line in Fig. \ref{fig:columnE}. We can see that this excellently captures the shape, normalisation and maximum of the simulation curves.  This validates our toy model and explains why the efficacy of wind driving  is such a strong function of energy.  In particular it demonstrates why $1000~\mathrm{eV}$ X-rays (employed by \citetalias{Wang_2017}) are too weakly interacting to heat material to $T_{\rm Bern}$, whereas energies of $500-700~\mathrm{eV}$ are most potent.
The fact that despite its simplicity, this model so well captures the dependence across a range of energies - which penetrate to depths with different densities and thus different associated cooling rates - suggests that it is largely the attenuation of radiation, and not the variation in cooling, that determines the base - this is in line with our earlier discussion of the assumption of optically-thin heating being insufficient.

% Comparison of columns between models (not energies)
Note that this model also explains the small differences between the Full and IRFLNoH models. We see in Fig. \ref{fig:efficiency} that the cooling rates only diverge as $T_{\rm Bern}$ becomes larger in the inner disc due to the reduced cooling of the IRFLNoH model which lowers $\epsilon_c$ (Fig. \ref{fig:efficiency}). This makes it easier to heat a larger column at small radii in the disc. Intermediate radii also see a somewhat increased potency for winds being driven by the harder frequencies (Fig. \ref{fig:columnE}), but the largest radii with the coldest $T_{\rm Bern}$ are essentially unaffected.

%If one wishes to take the simple approach of having few spectral bins, then a softer X-ray energy would be more representative, as we discuss further in Section ?.

\subsection{Effect of Luminosity}
\label{sec:highL}
\begin{figure*}
    \centering
    \includegraphics{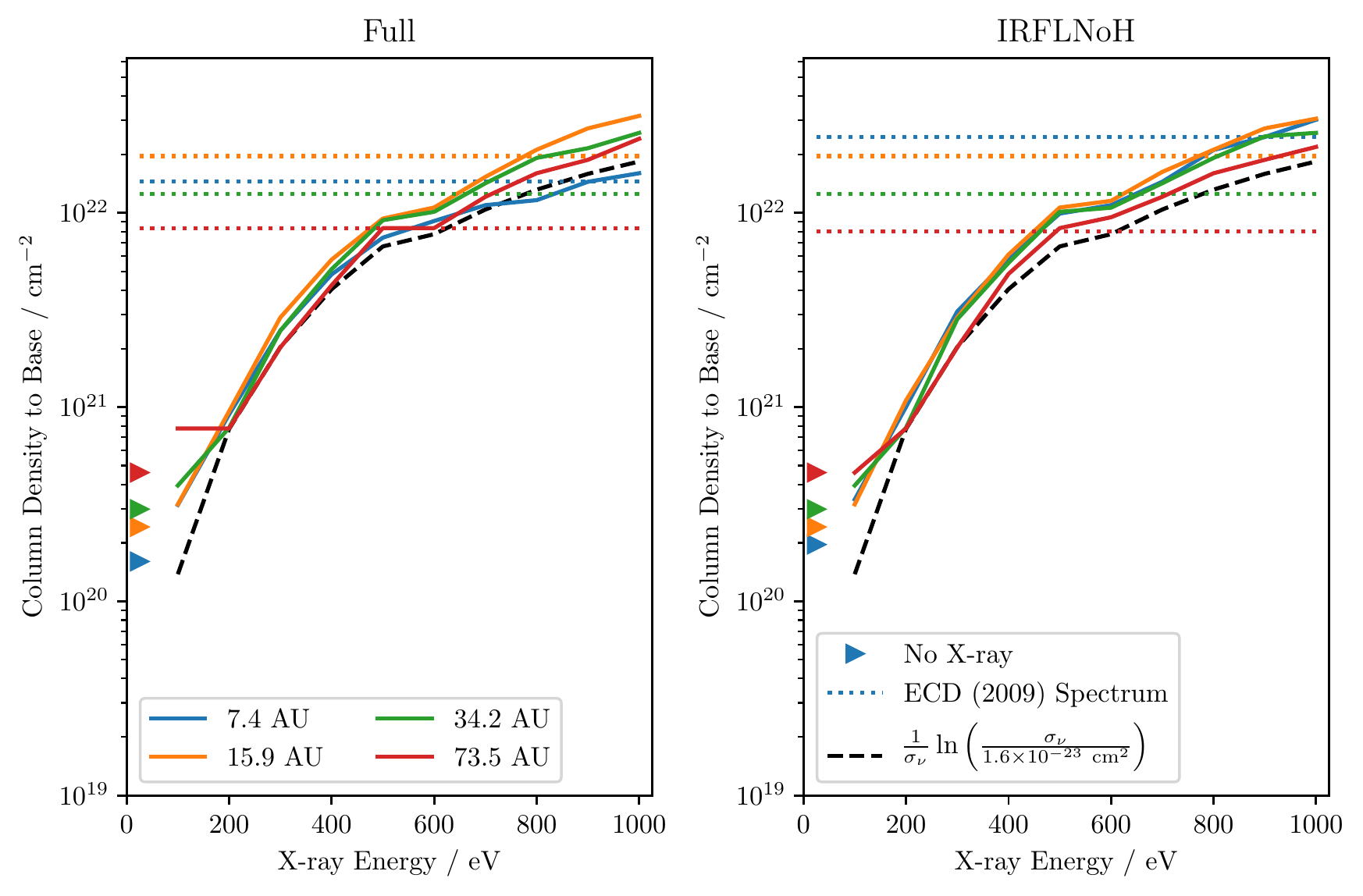}
    \caption{As Fig. 4 but for an increased X-ray luminosity $L_X= 1.6\times10^{31}~\mathrm{erg~s^{-1}}$.
    The coloured dotted lines are the values obtained using \citet{Ercolano_2009}'s spectrum.
    The dashed line is equation (\ref{eq:Neff}) for $\epsilon_c=1.6\times10^{-23}~\mathrm{cm^{2}}$.
    }
    \label{fig:columnE_highLX}
\end{figure*}

The spectrum of WG17 has relatively less X-ray compared to its UV flux (i.e. is overall softer) than that of \citep{Ercolano_2009}; therefore as well as the choice of the single X-ray band (the correct value of which will be determined by the shape of the X-ray spectrum), the relative luminosities could also be acting to diminish or enhance the role of X-ray between these studies.
Fig. \ref{fig:columnE_highLX} therefore shows the column density to the Bernoulli surface for each energy for our  X\#\#\# simulations which have $6.25$ times the X-ray luminosity of the S\#\#\# simulations and which therefore reproduce the ratio of $L_X/L_{\rm EUV}=0.8$ in the multifrequency input spectrum employed by \citet{Ercolano_2009}.

At the low energy end, where the optical depths are high, this has relatively little effect on the columns reached. Greater difference is seen as we move to higher energies, where the column no longer peaks around $500-700~\mathrm{eV}$; indeed among those frequencies  tested, $1000~\mathrm{eV}$ was the most effective. This is in line with our explanatory model - since $\epsilon_c \propto 1/\xi \propto 1/L_X$, then the appropriate $\epsilon_c\sim1.6\times10^{-23}$. With this lowered $\epsilon_c$ our model (black dashed line) remains an excellent fit and we would thus expect the highest column to be reached for X-ray  energy of around $1350~\mathrm{eV}$  in this case.

We note that the peak column $N\propto1/\epsilon_c \propto L_X$. Therefore, since we have argued it is reasonable to assume that the the column density in any wind scales with the mass loss rate of the wind, then one would expect $\dot{M} \propto L_X$; this is roughly as observed in previous hydrodynamical simulations \citep{Owen_2012,Picogna_2019} that found X-ray driven wind solutions.
Thus, by making more energy available in the X-ray, the ability of X-rays, in particular the harder bands, to drive a wind can be improved. Nevertheless, the choice of frequency still has a strong effect on the outcome. Since the X-ray luminosity is an inherently variable quantity (and closely tied to the stellar mass), the notion of the most effective energy for driving the wind will be linked to stellar properties.

Finally, we note that both the EUV luminosity and X-ray luminosity are likely to vary between stars. In models of EUV-driven winds under direct irradiation, the base densities (and wind densities) scale with the number of ionising photons as $n_{\rm base} \propto \Phi^{1/2}$ \citep[e.g.][]{Hollenbach_1994,Tanaka_2013} which can be understood from a simple Str\"{o}mgren-volume approach. The $\epsilon_c$ to be overcome in order for X-ray to heat below the EUV-heated base therefore also scales as $\Phi^{1/2}$; thus we could get the same result as here by lowering the EUV luminosity by two orders of magnitude.
Moreover if the X-ray luminosity scales more strongly than $\Phi^{1/2}$, then $\epsilon_c$ becomes a decreasing function of $L_X$ and so higher luminosity sources will be more likely to host X-ray-driven winds, while sufficiently low $L_X$ would lead to EUV-driven winds; these trends would be reversed if $L_X$ is a relatively weak function of $\Phi$.

\subsection{Monochromatic vs full-spectrum modelling}
\label{sec:full_spectrum}
A simplification in the above argument is that heating is assumed to be driven by only one frequency of radiation. In reality in these models, as aforementioned, FUV is assisting through photoionisation of carbon and sulfur\footnote{While the dust is also FUV heated and is a significant source of opacity, it is a net coolant of the gas in these regions.}, making it a little easier to launch a wind.

Moreover, a realistic  X-ray spectrum would have a range of bins with different efficiencies  - and with different individual luminosities and hence contributions to the spectrum - all working together: one might suppose that
such a spectrum would be intermediate in terms of heated column compared with monochromatic models at a few hundred and $1000$ eV since more energy is present in the effective bands than in the most extreme cases but it is not all concentrated there.
We use the spectrum of \cite{Ercolano_2009} as an illustrative example to explore this.
%Since frequencies $<1000~\mathrm{eV}$ are more efficient at heating the wind, we expect this to be better at creating an X-ray wind than a single bin at $1000~\mathrm{eV}$. From looking at the third column of Fig. \ref{fig:tempSpec}; this is clearly true of models E\_Full and E\_IRFLNoH - they can heat a larger column to escape than $1000~\mathrm{eV}$ alone.
%However, conversely the radiation is not centred around the most effective frequency but distributed more widely and hence we might expect the most effective frequency to provide an upper bound on the column that can be heated.

Fig. \ref{fig:columnE_highLX} shows, as dotted lines, the column achieved by Spectrum E at each radius.
As predicted, the heated column is somewhat intermediate between that at low energies and $1000~\mathrm{eV}$. % and spread between highly effective and less effective frequencies.
%\cathie{ However, in interpreting these results, it is important to realise that spectrum E has been normalised to the same EUV flux as \citetalias{Wang_2017} and corresponds to an order of  magnitude higher X-ray luminosity..... I agree with your comment: it's hard to try and disentangle the effects of a different spectrum and a different X-ray luminosity at the same time so I would advocate those further higher $L_X$ monochromatic runs...}
%In contrast with the above argument they are consistently higher, by a factor $\gtrsim3$, than that those achieved at the same radius by any single frequency.
Spectrum E heats a substantially higher column than any of the monochromatic spectra in our S\#\#\# series  (see Fig. \ref{fig:columnE}), mainly because of the roughly 10 fold higher total X-ray luminosity for models normalised to the same EUV flux.
%We note however, that due to the higher overall $L_X$, spectrum E heats a larger column than any frequency in our simulations with the S\#\#\# spectra.
Thus, the overall shape of the spectrum of \citet{Ercolano_2009}, namely that it is harder than \citetalias{Wang_2017}'s (and so when normalised to the same EUV flux has almost an order of magnitude higher X-ray luminosity) is a further key reason why it allows a somewhat deeper heating of a wind and leads to higher mass-loss rate, X-ray driven, models.

To quantify what we would expect for a continuous spectrum, we can attempt to generalise our model by replacing the single frequency treatment with an integral over frequency \citep[c.f. the attenuation factor of][]{Krolik_1983,Alexander_2004b}:
\begin{equation}
    L_X \sigma_\nu e^{-N\sigma_\nu} \to \int\limits_{E>100~\mathrm{eV}} \Phi_\nu \sigma_\nu e^{-N\sigma_\nu } d\nu
\end{equation}
for the spectral flux $\Phi_\nu$.
Thus our modified condition for sufficient heating to launch a wind becomes
\begin{equation}
    \epsilon_{\rm eff}(N) := \int\limits_{E>100~\mathrm{eV}} f_\nu \epsilon_\nu d\nu \geq \epsilon_c
    \label{eq:efficiencyeff}
\end{equation}
where $f_\nu = \Phi_\nu/L_X$ and thus the effective efficiency $\epsilon_{\rm eff}(N)$ is a flux-weighted average efficiency.
We can therefore iteratively calculate $\epsilon_{\rm eff}$ for increasingly large N until it no longer satisfies the inequality in equation (\ref{eq:efficiencyeff}); the maximum N will be our estimate of the heated column.

\begin{figure*}
    \centering
    \includegraphics{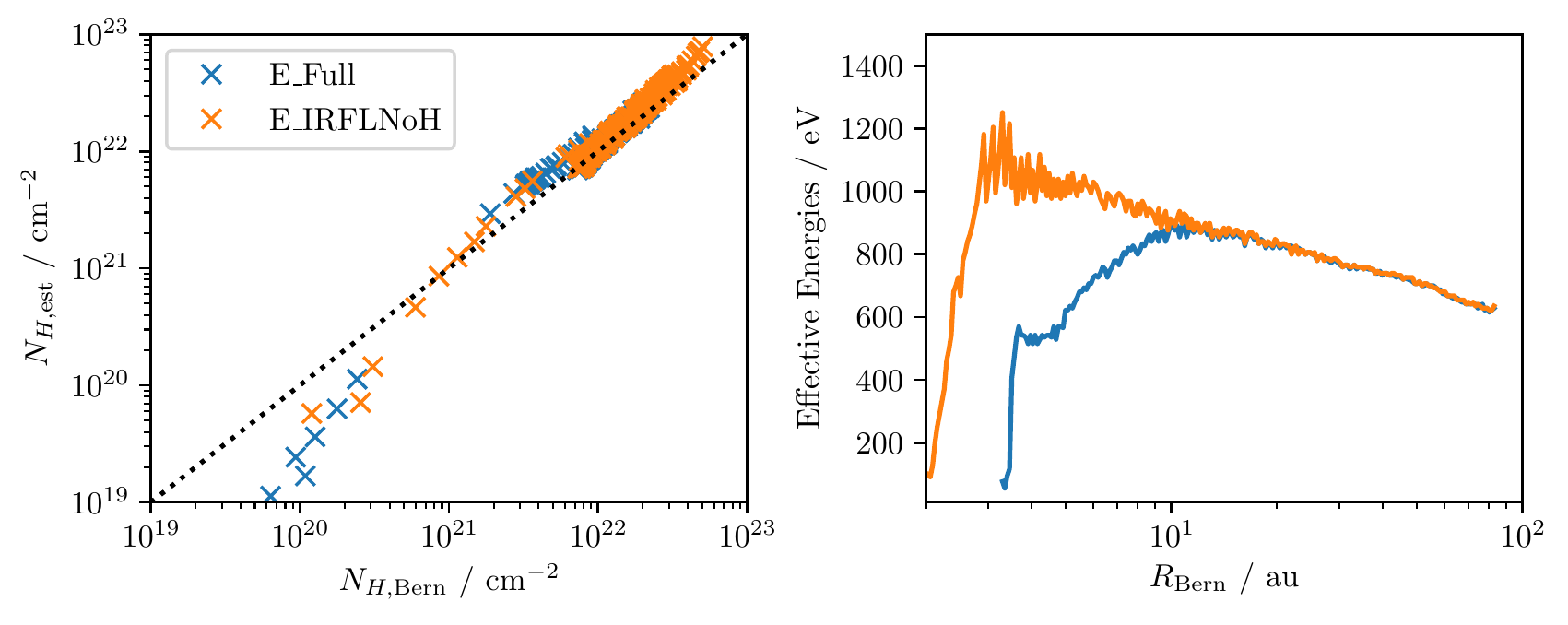}
    \caption{Left: the column to the Bernoulli surface estimated from the cooling rates and spectrum using the toy model as a function of the true column to Bernoulli surface in the simulations with spectrum E - good agreement is seen especially at large columns. Right: the lower of the two energies with heating efficiency equal to the effective efficiency of the whole spectrum at the relevant column as a function of radius.}
    \label{fig:continuousN}
\end{figure*}

The left-hand panel of Fig. \ref{fig:continuousN} shows the column density estimated from this method at each radius plotted against the column density to the Bernoulli surface for the E\_Full and E\_IRFLNoH models. There is a good agreement between the model and the true densities for most points at $\gtrsim10^{21}~\mathrm{cm^{-2}}$ and so we conclude that our model can be extended accurately to full spectra.

Clearly therefore, based on the arguments above, the most representative frequency is neither an ineffective one nor the most optimal one as much of the energy can be in less efficient bands.
However, for a given combination of $N$ and $\epsilon_{\rm eff}$, we can ask what single frequency would produce the same efficiency at that column. There are two solutions, the lower and high energy ones having cross sections $\sigma_1 = -\frac{1}{N} W_{-1}(-N\epsilon_{\rm eff})$ and $\sigma_2 = -\frac{1}{N} W_{0}(-N\epsilon_{\rm eff})$ where $W_0$ and $W_{-1}$ are the two real branches of the Lambert W function.
For each radius in the simulations with spectrum E, the lower of these two energies is indicated on the right-hand panel of Fig. \ref{fig:continuousN}, while the higher energy solution is generally not realistic for X-ray spectra of low-mass stars and so is not depicted.
As usual, very little difference is seen between the cooling models outside of $\sim10~\mathrm{au}$.
We expect these energies to be a function of column as higher columns will progressively attenuate the spectrum at its harder end, meaning that what reaches the base will be better approximated by softer energies.
Thus, as the column to the Bernoulli surface can vary with radius, the most representative energies also change, making it hard to reasonably pick a single frequency that would drive the wind everywhere with complete accuracy compared to a full spectrum.
Nevertheless, for both cooling models, outside the innermost few au, the appropriate energies are always $<1000~\mathrm{eV}$, further suggesting that this choice by \citetalias{Wang_2017} may not be an appropriate one. In the outer disc, the most representative energy is around $600~\mathrm{eV}$ and would expected to drive an X-ray wind from the whole disc given the cooling rates assumed here.

More recent spectra as used by \citet{Ercolano_2021} are somewhat softer than that used here, particularly for the lowest luminosity stars. These are therefore better represented by even softer energies from $400-800~\mathrm{eV}$ for $L_X=10^{31}~\mathrm{erg~s^{-1}}$ to close to $100~\mathrm{eV}$ for $L_X=10^{29}-10^{30}~\mathrm{erg~s^{-1}}$. This may lead to less effective photoevaporation as these energies are less effective than $\sim600~\mathrm{eV}$ due to their shallower penetration, though \cite{Ercolano_2021} do still see a substantial X-ray driven wind. Similarly, \citet{Nakatani_2018b} used the TW Hya spectrum from \citet{Nomura_2007} which is very soft due to its "soft X-ray excess" \citep{Gudel_2009}; its representative energies should therefore be very low, which may be a factor in their result that X-rays are ineffective drivers of photoevaporation on their own.

%Assuming a value of $\epsilon_c$, then for any given column $N\leq 1/(e\epsilon_c)$, a range of energies will be able to heat at least this column. This range is bounded by the X-rays with cross sections $\sigma_1 = \frac{1}{N} W_{-1}(-N\epsilon_c)$ and $\sigma_1 = \frac{1}{N} W_{0}(-N\epsilon_c)$ where $W_{-1}$ and $W_{0}$ are the two branches of the Lambert W function. As we increase $N$, the corresponding range of energies gets narrower, and the luminosity available to heat that column $L_{\rm eff}$ decreases. However, in order to determine $\epsilon_c$ we had to assume a corresponding X-ray luminosity $L_{X,c}$. Therefore, for any $\epsilon_c$, there is a maximum $N$ such that the effective X-ray luminosity $L_{\rm eff}\geqL_{X,c}$. Ranging over all possible $L_{X,c} \leq L_X$, the maximum value of N becomes our estimate for the maximum heated column. 

%Applying this procedure to models E\_Full and E\_IRFLNoH at $73~\mathrm{au}$, we see that the most effective range is $167-298~\mathrm{eV}$, covering a luminosity of around $1.2\times10^{30}\mathrm{erg~s^{-1}}$. However, the corresponding column of $5\times10^{20}~\mathrm{cm^{-2}}$ is nearly an order of magnitude too low. This is because we effectively assumed a top hat which is clearly an underestimate

\section{Discussion: Cooling and Heating processes}
\label{sec:cooling}

\begin{figure*}
    \centering
    \includegraphics{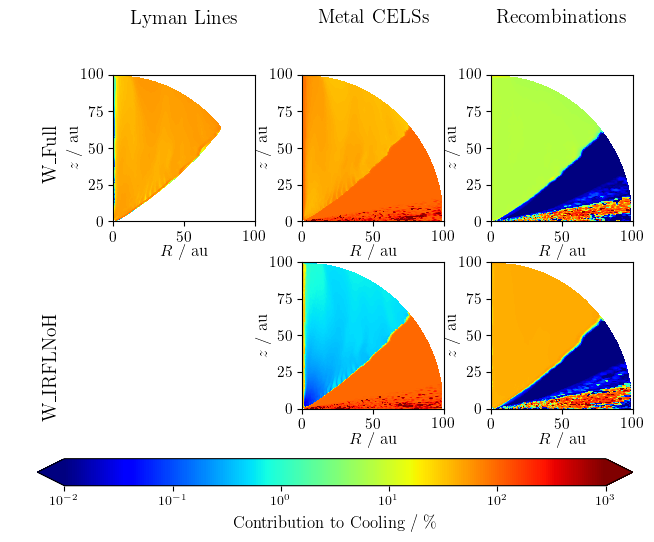}
    \caption{The percentage contribution to the cooling from 3 key processes: permitted line Lyman radiation from collisionally excited H, forbidden line radiation from collisionally excited metals and recombinations in the fiducial simulations. The top row shows the Full cooling model where cooling is dominated by Lyman lines and metal CELs. The bottom row shows the IRFLNoH model - here Lyman lines are switched off and the Metal CELs severely suppressed, increasing the role of recombinations in the wind region.
    Note that percentages greater than 100 are recorded near the midplane as \textsc{mocassin} treats dust as a coolant but here it can become warmer than the gas and has a net heating effect i.e. a negative cooling contribution.}
    \label{fig:coolFiducial}
\end{figure*}
We have so far explored how the ability of X-rays to heat the gas sufficiently depends on their frequency and shown that only those with energies of a few $100~\mathrm{eV}$ can overcome the cooling included in our radiative transfer models. It is therefore important to consider more closely the impact of differences in how cooling is treated between photoevaporation models.
To provide a baseline comparison, we first determine the dominant cooling channels in our models, before exploring the impact of the additional cooling channels discussed by \citetalias{Wang_2017} as differences in methodology between their work and that of \citet{Owen_2010}: neutral sulfur, adiabatic cooling and molecular cooling.

Fig. \ref{fig:coolFiducial} shows three key cooling channels in our models - from left to right: collisionally exited Lyman lines of H, collisionally excited forbidden lines of metals and recombinations - for the fiducial spectrum W. The top row indicates their fractional contribution to the cooling for the Full cooling and the bottom row for IRFLNoH.

For the Full cooling, the wind has fairly equal contributions to the cooling from Lyman radiation and metal CELs, with the former dominating slightly at larger radii and vice versa. Metal CELs are almost entirely responsible for cooling below the wind base, while recombinations play only a minor role in the wind and none below the base where the material is most neutral.
For the IRFLNoH cooling, the Lyman lines have been switched off and play no part in the cooling. The metal CELs are still dominant below the base, but are heavily suppressed in the wind region as this contribution was largely down to optical lines, particularly those of \ion{S}{II}. Instead, cooling in the wind is now almost entirely dominated by recombinations, which was the only significant non-adiabatic cooling in this region according to \citetalias{Wang_2017}.

This confirms that in the bulk of the wind, we expect significantly higher levels of non-adiabatic cooling than found by \citetalias{Wang_2017} in the form of the Lyman lines and optical CELs. This is sufficient to explain the cooler wind temperatures seen for the Full cooling model and indeed in most photoevaporation models \citep[e.g.][]{Owen_2012}.
The reason \citet{Wang_2017} do not see significant Lyman cooling is because the optical depth of these lines is approximately $10^3 \lesssim \tau_{\rm Ly\alpha} \lesssim 10^5$ with escape probabilities $10^{-5}\to10^{-4}$ meaning this cooling is several orders of magnitude weaker if the radiation is not allowed to escape through the line wings or as a result of absorption and re-radiation in the IR by dust.

On the other hand, at the temperatures around $T_{\rm Bern}$ and below the wind base, there is less difference between the cooling models. This region is well-described as a PDR - which are typically cooled mainly by [\ion{O}{I}] $63~\mathrm{\mu m}$ and [\ion{C}{II}] $158~\mathrm{\mu m}$ \citep{Tielens_1985} - and so the set of coolants considered by \citet{Wang_2017} is appropriate here.

\subsection{Sulfur}
Unlike previous works applying \textsc{mocassin}, our models include neutral sulfur. Within the wind itself, sulfur is mostly doubly ionised, with a non negligible contribution from singly ionised sulfur but little neutral sulfur remaining. On the other hand, below the base sulfur is mostly singly ionised by FUV in the heated region, transitioning to mostly neutral beyond this. The inclusion of neutral sulfur will therefore make negligible difference to the temperatures of the wind itself, while having the most impact wherever the disc becomes optically thin to FUV; the X-rays most efficient at heating this region and thus those most affected will be those with similar cross sections to FUV, i.e. with energies $\gtrsim 1~\mathrm{keV}$. However, since our results show that X-rays (including the hard $1~\mathrm{keV}$ X-rays for some radii) are able to heat material to escape even with the inclusion of neutral sulfur, we must conclude that it is not a critical cause for the differences seen between \citetalias{Wang_2017} and \citet{Owen_2010}. On the other hand, there are optical CELs of \ion{S}{II} that are significant coolants and are not included by \citetalias{Wang_2017}, which is one contributing factor to their hot wind temperatures.

\subsection{Hydrodynamical cooling}
The main difference that \citet{Wang_2017} claim in the wind region is that adiabatic expansion overwhelmingly dominates the cooling and offsets the photoionisation heating. We have already seen how the inclusion of optical forbidden lines and unattenuated Lyman lines can increase the non-adiabatic cooling budget significantly, making it more competitive with the adiabatic contribution. Before quantifying how much so, it is useful to discuss what is meant by adiabatic cooling.

The evolution of the total energy density $E = \rho\epsilon = \frac{1}{2}\rho v^2 + \rho u + \rho \Phi$ in a static potential $\Phi$ is described by the energy equation
\begin{equation}
    \frac{\partial E}{\partial t} + \nabla\cdot((E+P)\vec{v}) = \rho(\Gamma-\Lambda)
    \label{eq:energy}
    ,
\end{equation}
for heating and cooling rates per unit mass $\Gamma$ and $\Lambda$ respectively.
%This can be broken down into
%In the steady state, assuming thermal equilibrium 

The corresponding equation (in conservative form) for the thermal energy density $\rho u$ is
\begin{align}
    % \frac{D}{Dt}\left( \frac{1}{2}\rho v^2 \right) &= \frac{\partial}{\partial t}\left( \frac{1}{2}\rho v^2 \right) + \nabla\cdot\left( \frac{1}{2}\rho v^2 \vec{v}\right) = -\rho \vec{v}\cdot\nabla\Phi -\vec{v}\cdot\nabla P \\
    \frac{\partial}{\partial t}\left( \rho u \right) + \nabla\cdot\left( \rho u \vec{v} \right)
    = \rho \frac{Du}{Dt}
    = \rho(\Gamma-\Lambda) - P \nabla\cdot\vec{v}
    \label{eq:thermalenergy}
    .
\end{align}
The additional term on the right hand side compared to equation (\ref{eq:energy}) represents the "PdV" work done on fluid element by expansion in the presence of a diverging velocity field which adiabatically cools the gas. The energy lost from the thermal contribution is used to accelerate the wind by pressure gradients along the streamlines.
%The last term of the first equation provides a method for the same pressure gradients that set the diverging velocity field to instead accelerate the gas and loosely speaking convert the thermal energy to kinetic energy in an enthalpy-conserving process.

However, in establishing a steady state thermal balance, we are more interested in the thermal evolution at a particular location:
\begin{equation}
    \frac{\partial u}{\partial t}
    = (\Gamma-\Lambda) - (\gamma-1) u \nabla \cdot \vec{v} - \vec{v} \cdot \nabla u
    .
\end{equation}
Thus, while adiabatic cooling is relevant for the cooling of a fluid element, the advection of thermal energy also plays an important role when setting the thermal balance in an Eulerian sense.
This thermal flux could potentially offset the adiabatic cooling if material flows from hot to cold and should also be considered.

%The second equation may be expanded to give the Lagrangian derivative of the thermal energy per unit mass
%\begin{equation}
%    \frac{D\epsilon}{Dt} = (\Gamma-\Lambda) - \frac{P}{\rho} \nabla\cdot\vec{v} = (\Gamma-\Lambda) - (\gamma-1)\epsilon \nabla\cdot\vec{v}
%\end{equation}
%or, per particle, where $\rho = \mu m_H n$ and $\epsilon_n = \mu m_H \epsilon$
%\begin{equation}
%    \frac{D\epsilon_n}{Dt} = (\Gamma_n-\Lambda_n) - kT \nabla\cdot\vec{v}
%\end{equation}
%Thus, the adiabatic cooling per particle experienced by a parcel of gas due to pdV work is $kT \nabla\cdot\vec{v}$.
%However for a particular grid cell (i.e. in an Eulerian sense), we must consider that the former version becomes
%\begin{equation}
%    \frac{\partial\epsilon}{\partial t} = (\Gamma-\Lambda) - \frac{P}{\rho} \nabla\cdot\vec{v} - \vec{v}\cdot\nabla\epsilon
%\end{equation}
%or
%\begin{equation}
%    \frac{\partial\epsilon_n}{\partial t} = (\Gamma_n-\Lambda_n) - kT \nabla\cdot\vec{v} - \frac{k}{\gamma-1} \vec{v}\cdot\nabla(T)
%\end{equation}
%i.e. the adiabatic cooling may be offset by advection of thermal energy.

In steady state, integrating equation (\ref{eq:energy}) over a volume following a streamline bundle and using mass conservation gives
\begin{equation}
    \dot{M} \Delta \epsilon_{\rm tot} = L_{\rm heat} - L_{\rm cool}
    ,
\end{equation}
where $\Delta \epsilon$ is the difference between the mass-flux-weighted average energy density at either end of the bundle.
\citet{Owen_2010} argue that since $\dot{M}\Delta\epsilon_{\rm tot} \lesssim 8\% L_X$, then the advected energy is negligible compared to $L_{\rm heat}$ can assume $L_{\rm heat} \approx L_{\rm cool}$. This relies on the assumption that $L_{\rm heat}\approx L_X$, which may not be true if significant luminosities lie at harder energies that penetrate through the wind base and are absorbed at longer columns. While this result has not been thoroughly investigated for a range of luminosities, since $\dot{M}$ is typically found to scale approximately linearly with $L_X$ \citep{Owen_2011,Picogna_2019} - except at the highest luminosities -  then the argument should translate. Moreover, in all the spectra here, $L_{\rm heat} \gtrsim L_{EUV} \gtrsim L_X$ so in any case this wouldn't affect the conclusion strongly.

Moreover, integrating for the thermal energy density $u$,
\begin{equation}
    \dot{M} \Delta u_{\rm tot} = L_{\rm heat} - L_{\rm cool} - L_{\rm adiabatic}
    .
\end{equation}
Hence, by comparison we conclude that $\dot{M} \Delta \epsilon_{\rm tot} = \dot{M} \Delta u_{\rm tot} + L_{\rm adiabatic}$ and the advected energy is the net result of any advected thermal energy less any adiabatic cooling. Since the wind consists of unbound material, and is being accelerated, it is reasonable to assume the dominant contribution to the advected energy is an increase in kinetic energy - since the wind ends up supersonic, this is likely of greater magnitude than any change in thermal energy i.e. $\dot{M} \Delta \epsilon_{\rm tot} >> \dot{M} \Delta u_{\rm tot}$ and hence $L_{\rm adiabatic} \approx \dot{M} \Delta \epsilon_{\rm tot}$ and probably shouldn't be significantly offset by thermal advection.

%\begin{itemize}
%    \item Firstly, that $L_{\rm heat} \gtrsim L_X$ in the wind. Since $\Delta\epsilon$ is measured from launch to edge of grid, we should not include any heating below the wind base: \citetalias{Wang_2017} would suggest that below the base is where the X-ray heating is important, though $L_{\rm heat} \gtrsim L_{EUV} \gtrsim L_X$ so in any case this wouldn't affect the conclusion strongly.
%    \item Secondly that $L_{\rm adiabatic} \sim \dot{M}\Delta\epsilon_{\rm tot}$. If the energy is dominated by thermal energy everywhere, then this may be reasonable as the thermal flux and adiabatic cooling rely on the same quantities, though combined differently - for example in an isothermal case, there should be no net advection of thermal energy. However at high Mach numbers, most of the wind's energy is in kinetic energy rather than in thermal energy as at the base. Therefore a modest increase in the energy flux of the wind may require a more substantial drop in the thermal energy, which in turn could result from significant adiabatic cooling. 
%\end{itemize}

\begin{figure*}
    \centering
    \includegraphics{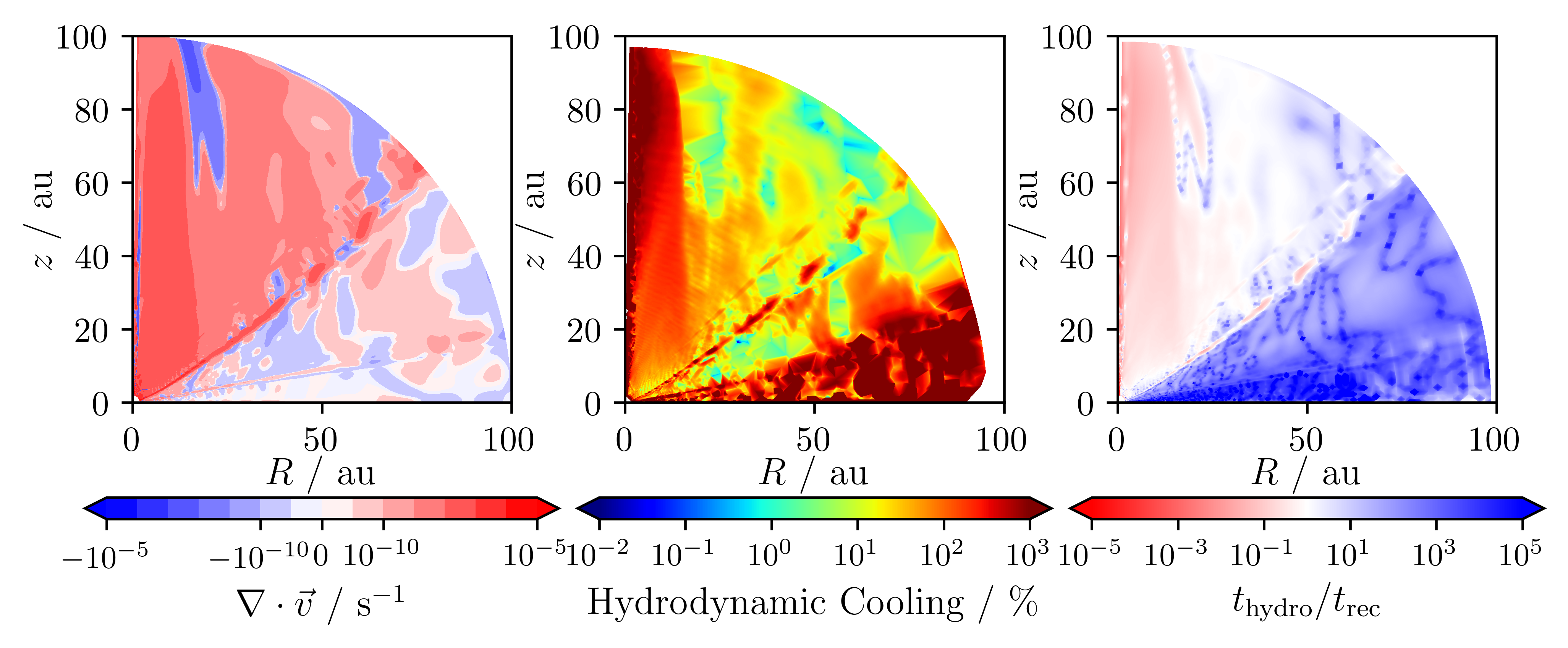}
    \caption{Estimates of the significance of hydrodynamical cooling for simulation W\_Full. The leftmost panel shows the divergence of the velocity grid from \citetalias{Wang_2017} with diverging flows in red and converging flows in blue. The central panel shows the hydrodynamical cooling relative to the total \textsc{mocassin} cooling expressed as a percentage. The large values near the midplane are likely an artefact of the low cooling rates there. The rightmost panel shows an estimate of the ratio between the hydrodynamic and cooling (recombination) timescales with red regions indicating shorter hydrodynamic timescales and blue indicating regions where the radiative equilibrium is reasonable to assume.}
    \label{fig:divv}
\end{figure*}

Fig. \ref{fig:divv} shows quantities relevant to hydrodynamical cooling for the W\_Full simulation. Firstly, the left-most panel shows the divergence of the velocity field $\nabla\cdot\vec{v}$ from \citetalias{Wang_2017} (since we do not recalculate this for our temperature field). Indeed over most of the wind volume, the velocity field is diverging which would result in cooling of the material. It can be seen that this is particularly strong in a column at $R\lesssim20~\mathrm{au}$ - this is a result of a strong acceleration in the radial direction.

The net hydrodynamical cooling as a percentage of the non-adiabatic calculation from \textsc{mocassin} (using the temperatures and cooling rates of the W\_Full simulation) is shown in the middle panel of Fig. \ref{fig:divv}. We can see that correspondingly, while in most of the the volume, it can only account for around $10\%$ of the cooling \citep[i.e. similar to the value found by][for material originating at $R\approx20~\mathrm{au}$]{Owen_2010} - making it a not insignificant (compared to e.g. recombinations) but nevertheless non-dominant contribution - adiabatic cooling is important at $R\lesssim20~\mathrm{au}$\footnote{While included in the calculation, thermal advection is negligible here as the wind region is close to isothermal.}. This suggests that the cooler temperatures seen by \citet{Wang_2017} in this region than in our W\_IRFLNoH simulation are a result of additional adiabatic cooling, which is strongest - at least for a mild temperature gradient - at small radii.
There are also a few hotspots where adiabatic cooling may be important near the base of the wind as the material is accelerated through the wind base suggesting adiabatic cooling may have some effect on the launching of the wind.

As a further check, in the right hand panel we show the ratio of the hydrodynamical timescale (estimated as $|\nabla\cdot\vec{v}|^{-1}$) and the recombination timescale ($1.5\times10^9T_e^{0.8}n_e^{-1}$, which is usually the longest microphysical timescale \citet{Ferland_1979,Salz_2015}). Again, we see that in the bulk of the wind and base the hydrodynamical timescale is around an order of magnitude longer and we can safely assume radiative thermal equilibrium but that near the z-axis the hydrodynamical timescale is shorter and the assumption may break down.

It is notable that the estimates of timescales are much more comparable than were found by \citet{Picogna_2019}. On the one hand, the temperatures are a little higher here which increases the typical velocity scale ($c_S\propto T^{0.5}$) and hence decreases the hydrodynamical timescale, while the recombination timescale increases as the electrons are more energetic and harder to recapture. Moreover, the hydrodynamical timescale is independent of density, while the timescales of two-body non-adiabatic cooling processes are longer at the low densities of the EUV-driven density profile of \citetalias{Wang_2017} compared to the higher densities in \citep{Picogna_2019}'s X-ray driven wind (although lower X-ray luminosities may drive somewhat less dense winds in which this timescale is not so long and hence radiative equilibrium a less robust assumption).

We conclude that the contribution from adiabatic cooling shown in Fig.  \ref{fig:divv} probably represents an upper bound; this contribution should be less significant in the cooler, denser X-ray driven winds which we argue should result from the use of a softer X-ray spectrum than that employed by \citetalias{Wang_2017}. Nevertheless, adiabatic cooling should probably be considered further across the X-ray luminosity range, particularly when modelling the inner wind regions and their tracers \citep[e.g. \ion{O}{I} $6300~\text{\r{A}}$][]{Ercolano_2016}.
 
\subsection{Molecules}
\label{sec:molecules}
%\cathie{Comment: I don't disagree with anything you are saying here but I wonder whether it's worth a section? ....tbd...}
%\textit{I've attempted a bit of a rewrite of this section to make it a little more concise but a little more quantitative too to justify itself - let me know if it's an improvement?}

Molecules, likely to be present in the underlying disc, are the final missing piece of our model compared to that of \citetalias{Wang_2017} or \citet{Nakatani_2018b}. We thus conclude our exploration of the different cooling contributions by considering the potential consequences of including molecules on ability of X-rays to launch a wind and the resultant wind mass loss rates.

In Section \ref{sec:fiducial} we showed that irradiating \citetalias{Wang_2017}'s density grids using \textsc{mocassin} produced warmer temperatures below the IF and in Section \ref{sec:varybands} that this persisted once X-rays were removed entirely.
%For simulations U\_Full and U\_IRFLNoH, most of the heating below the base comes from FUV photoionisation of metals, especially carbon (and to a lesser extent sulfur) so by including molecules, one could lock up the C in CO, reducing the ability of photionisation to heat this region. Otherwise,
This implies extra cooling is needed below the base to fully reproduce \citet{Wang_2017}: since sulfur is included in our models and adiabatic cooling is relatively negligible in this region the best candidate is molecular cooling: rovibrational lines of molecular species (partiularly $\mathrm{H_2}$, $\mathrm{H_2O}$, OH) can be the dominant radiative processes just below the base \citepalias{Wang_2017}.

By increasing the available cooling at temperatures $\sim T_{\rm Bern}$, further cooling from molecules would affect the quantitative results of Section \ref{sec:toymodel} as to both a) which X-rays can launch a wind and b) which are most effective at doing so.
These mechanisms could be particularly significant in the cooler outer disc or at high optical depths to FUV, where molecular survival is more likely.
However, even at small/intermediate radii, extra cooling should reduce the ability of $1000~\mathrm{eV}$ X-ray to drive a wind somewhat (cf the Full models versus the IRFLNoH models) potentially even to the extent of eliminating X-ray driven winds entirely \citepalias[as seen by][]{Wang_2017}.

Properly quantifying the contribution of molecules is beyond the scope of this work, but since the additional cooling would increase $\epsilon_c$ (equation \ref{eq:efficiencyc}), then we can start by considering the impact of some representative increases in this parameter on the column which each X-ray frequency can heat (equation \ref{eq:Neff}) and which frequency is optimal for launching a wind.
%In particular, we can quantify how much of an increase would completely eliminate the ability of X-rays to launch a wind.

\begin{figure}
    \centering
    \includegraphics[width=\linewidth]{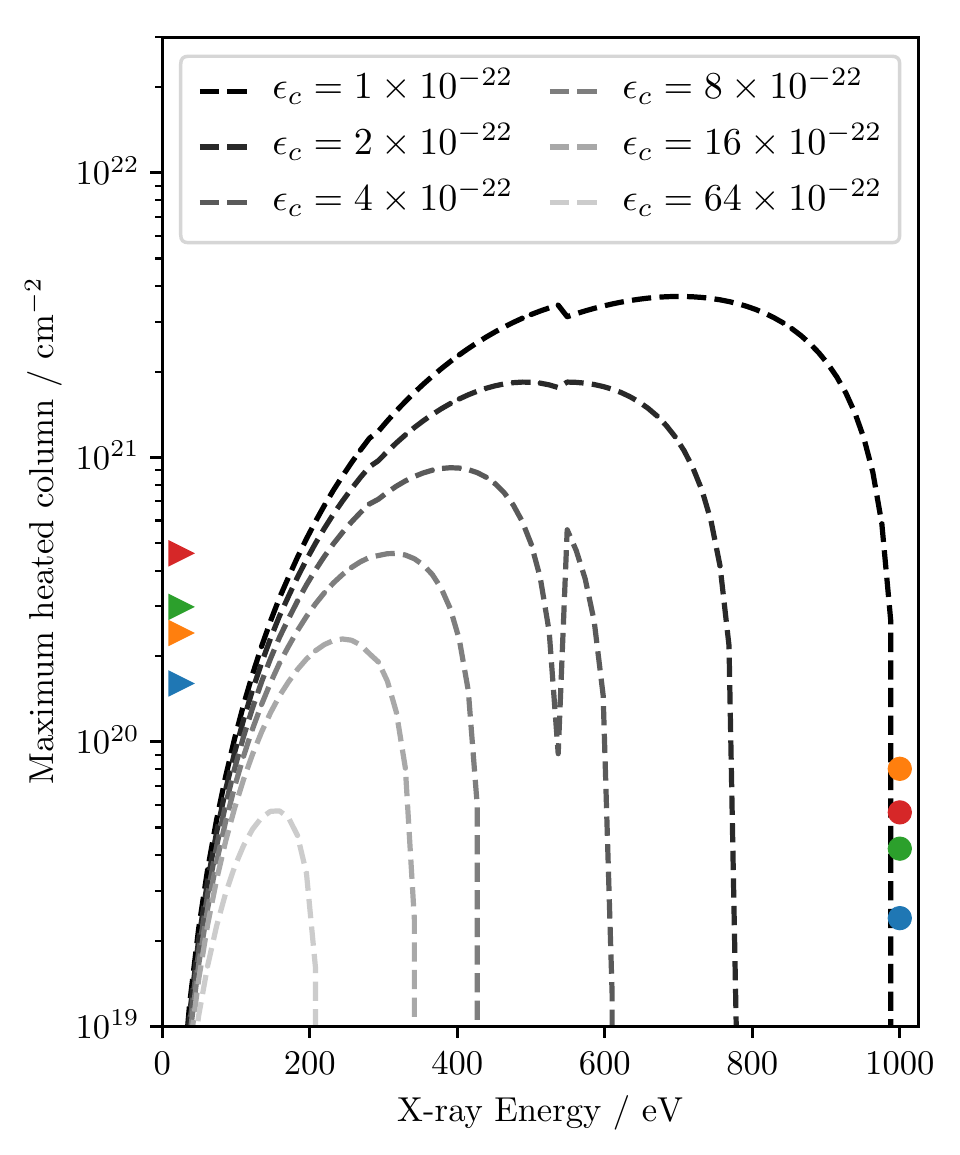}
    \caption{Effect of additional cooling on the maximum column that can be heated to $T_{\rm Bern}$. Additional cooling is parametrized as an increase in $\epsilon_c$ and represented by increasingly light colours. The kinks in some curves are because of a non-monotonicity in the photoionisation cross-section due to inner shell ionisation of oxygen. The coloured triangles represent the column reached in our UV-only model for the same range of radii as in \ref{fig:columnE}.}
    \label{fig:columnE_eps}
\end{figure}
Fig. \ref{fig:columnE_eps} shows that the curve of maximum heated column vs energy shifts down to lower columns (and to the left, peaking at lower energies) as $\epsilon_c$ is increased. An increase by a factor $\sim 8$ would be needed to prevent any X-ray from being able to heat a higher column than our UV-only simulations to $T_{\rm Bern}$ at large radii, and an increase by $\gtrsim 16$ would be needed to achieve this at all radii. Moreover, in this limit, only very soft X-rays $200-400$ would have any significant heating effect.
However, this is likely an underestimate of the necessary cooling since the column heated in these reference simulations would likely also be reduced somewhat towards the values found by \citetalias{Wang_2017}.
An increase in $\epsilon_c$ of more like $30-100\times$ may therefore be required to prevent any single X-ray from heating the wind.

%Additional cooling may therefore actually need to be a factor of  more than included in \textsc{mocassin} to prevent any single X-ray frequency from heating the wind. 
%launch a wind at large radii (although $\epsilon_c$ may already be a factor $\sim 2$ higher there due to geometric dilution), and an increase
Therefore to significantly affect our conclusion about the viability of X-ray wind launching at softer energies, molecules - or any other additional cooling not included in \textsc{mocassin} - would need to contribute at least an order of magnitude more cooling than the atomic processes here modelled. It is worth noting however, that such an increase in $\epsilon_c$ could also be caused by a decrease in the X-ray luminosity available to heat the wind, or the efficiency with which X-rays are able to deposit the absorbed energy into the gas due to losses to secondary ionisation by the photoelectrons before they thermalise (see Section \ref{sec:f_X}). Conversely, increases in these parameters would make it harder for molecular cooling to prevent an X-ray heated wind.

\begin{figure*}
    \centering
    \includegraphics[width=\linewidth]{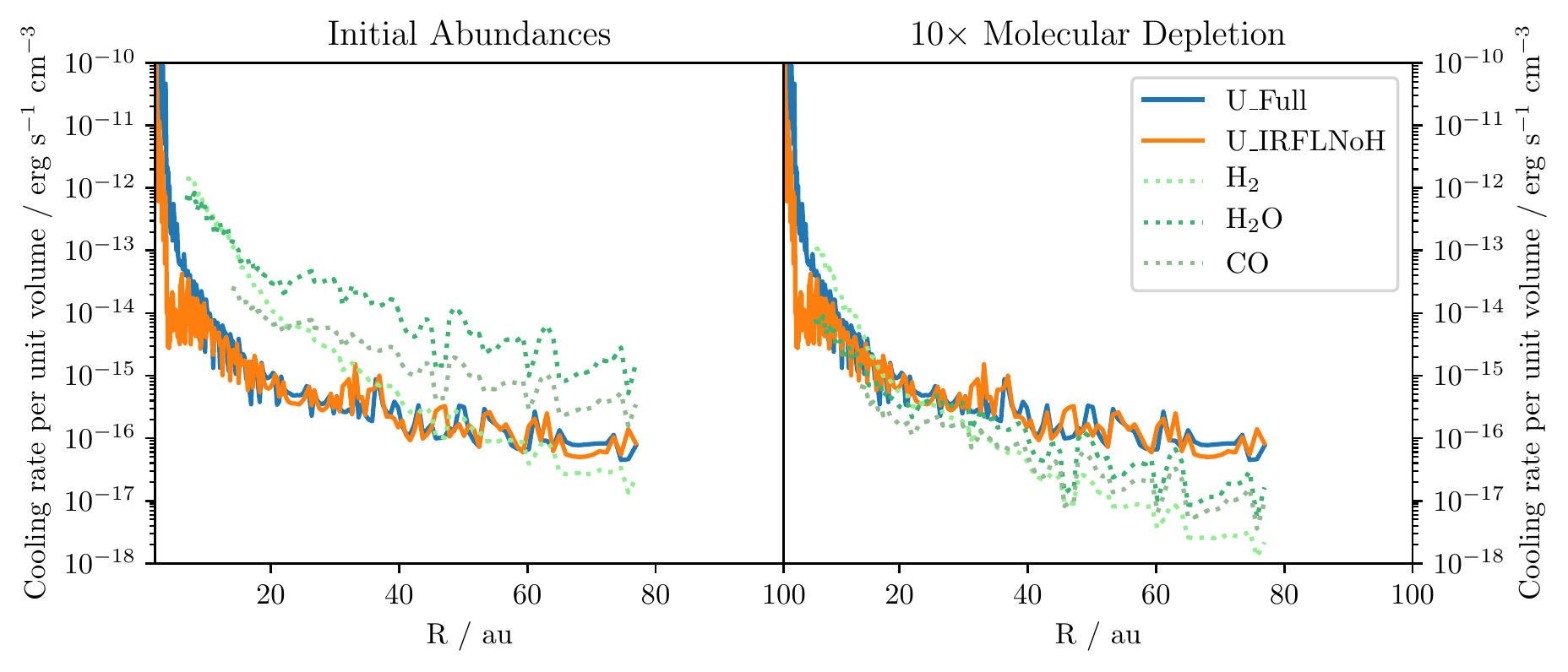}
    \caption{Cooling rates per unit volume for the U\_Full (blue) and U\_IRFLNoH (orange) simulations at the Bernoulli surface. The dotted lines show the possible contribution of additional molecular cooling from $\mathrm{H_2}$, $\mathrm{H_2O}$ and  $\mathrm{CO}$, on the left-hand panel with maximal abundances assuming all atoms are in molecules and on the right-hand panel with molecular abundances depleted by 10 times from the maximal values.}
    \label{fig:molCool}
\end{figure*}

To estimate whether molecules have this potential, we provide estimates for the impact of molecular ro-vibrational cooling along the Bernoulli surface from the U\_IRFLNoH model using the tabulations for $\mathrm{H_2}$, $\mathrm{H_2O}$ and $\mathrm{CO}$ from \citet{Neufeld_1993}.
We choose to match the initial molecular abundances of \citetalias{Wang_2017} and in each case we assume that the wind base is optically thin and so set the optical depth parameter to its minimal tabulated value. The calculated cooling rates are shown alongside those from the U\_Full and U\_IRFLNoH simulations in the left hand panel of Fig. \ref{fig:molCool}. The most significant cooling typically comes from water, which under these assumptions can contribute more than an order of magnitude more cooling than the atomic processes thus suggesting a potentially important role for molecular cooling in the framework set out above.

However, there are two important caveats: firstly, that we ignore any possible molecular heating. For example FUV pumping into the Lyman and Werner bands followed by collisional de-excitation from the vibrationally excited states of the ground electronic state can result in net heating of gas by $\mathrm{H_2}$.
Secondly, in following \citetalias{Wang_2017}'s inital molecular abundances we have made the most generous assumption that all C, O and H will be in molecules. In reality molecular abundances are likely to be somewhat lower than this at the base: since FUV is generally more penetrating than EUV or soft X-rays, the base will be optically thin to FUV which will lead to molecular dissociation. Moreover, the warm upper layers of protoplanetary discs are frequently observed to be depleted in volatile molecules such as $\mathrm{H_2O}$ by a couple of orders of magnitude \citep[e.g.][]{Du_2017} due to freeze-out onto ice grains that settle to the midplane (though such processes would also affect atomic coolants).
The right hand panel of \ref{fig:molCool} demonstrates that an order of magnitude depletion in all molecular abundances near the base would be sufficient to make them subdominant coolants to atoms.

A further caveat is that we have only considered molecular effects in the context of a single X-ray frequency. As discussed in Section \ref{sec:full_spectrum}, when a full spectrum is considered the heated column is somewhat less than for the most efficient frequencies. Our estimate that an order-of-magnitude more cooling may be needed to prevent an X-ray wind may therefore be a slight overestimate and molecules may not need to prevent all X-ray energies from heating the wind. However a moderate increase of around a factor 4 can likely still be tolerated, since in Section \ref{sec:full_spectrum} we showed that $\sim600~\mathrm{eV}$ would be representative of the integrated spectrum and effective heating at $\gtrsim600~\mathrm{eV}$ is prevented for a factor $\gtrsim4$ increase in cooling. Other spectra used in the recent literature \citep{Nomura_2007,Ercolano_2021} are softer than those used here so the representative energies are in regions that are more robust against being rendered ineffective by additional cooling since they lie in the attenuation-limited (optically thick) regime (whereas additional cooling progressively limits the effect of harder energies).
However as aforementioned, the effects of additional cooling and the X-ray luminosity are degenerate, so since the hardest spectra tend to be found for stars with the highest $L_X$, the X-ray heating for harder spectra has a `head start' against the effects of additional molecular cooling.
%Therefore molecules may not need to be strong enough to prevent the softer frequencies from heating if sufficiently little luminosity is incident at the frequencies that remain effective.

%equivalent to assuming that all carbon is contained in $\mathrm{CO}$, any remaining oxygen in $\mathrm{H_2O}$ and finally any remaining hydrogen in $\mathrm{H_2}$.

%This can explain the difference in the column to the Bernoulli surface between our simulations and the temperature profile from \citetalias{Wang_2017}.

Our simulations show hotter temperatures below the base.
The thermal structure here is also important to determining the mass loss rates as it determines hydrostatic equilibrium.
For example
%A secondary consequence of our hotter region below the base is that this region is not in hydrostatic equilibrium in our simulations but has overly high pressures.
with only the cooling present in our simulations, one would therefore expect to see the disc region puff up more, which can result in the disc intercepting more of the driving radiation, assisting mass loss rates \citepalias{Wang_2017}.
Furthermore, the momentum flux (and hence mass loss rate) of the wind depends on the pressure jump across the base, which depends on the temperature to which the underlying material is heated.

%Beyond affecting how the deeply the X-rays can heat the disc, decreasing the thickness of the disc is thus a further way that molecular cooling works to suppress heating and mass loss rates.

%Furthermore, even for the same density structure, mass loss rates would be lower for a cooler underlying disc.
%By applying continuity of mass and momentum and eliminating the density on the wind side, one can estimate the mass loss rate (per unit area) in the wind in terms of the Mach number at the wind base $\mathcal{M}_{\rm b}$ and underlying disc pressure $P_{\rm th, disc}$ and sound speed $c_{\rm S, disc}$ as
%\begin{equation}
%    \dot{M} \approx \frac{\mathcal{M}_{\rm b}}{1+\mathcal{M}_{\rm %b}^2} \frac{P_{\rm th, disc}}{c_{\rm S, disc}}
%    .
%\end{equation}
Thus, as an upper bound, molecules certainly would lead to X-rays being able to unbind gas less dense gas (at lower columns) than otherwise and thus decrease mass loss rates.
Moreover, without molecules we likely overestimate the temperature/pressure on the underside of the disc somewhat affecting its hydrostatic structure and momentum flux, which would lead to an overestimate of mass loss rates if we tried to infer them from our models.
This is all consistent with the fact that \citetalias{Wang_2017}'s fiducial model has a lower mass loss rate than their OECA10 analog which did not include molecular cooling.
Moreover, like \citetalias{Wang_2017}, \citet{Nakatani_2018b}, who also included molecules, found that X-rays were not able to drive a wind, despite their softer spectrum that peaks at a few $100~\mathrm{eV}$. However, they do not include cooling from water, which seems to be the most significant molecular coolant if abundant.

Altogether it is unclear whether, other than in the most generous scenario, molecular cooling could reduce the role of X-ray enough to result in an EUV-driven wind. A self-consistent calculation of molecular abundances with a realistic X-ray spectrum is needed to more accurately determine their role in competing against X-ray heating.

\subsection{X-ray Heating Fraction}
\label{sec:f_X}

In applying our explanatory model, we have set $f_X=1$ as is appropriate for a UV-heated wind. As aforementioned, the true value may be lower, which is likely the case at the base of an X-ray driven wind since X-ray heated material has low levels of ionisation. Given the degeneracy noted earlier, the effect of such a lower $f_X$ can be understood in exactly the same terms as any additional cooling, by using Fig. \ref{fig:columnE_eps}. That fact that the explanatory model with $f_X=1$ still fits the data well however, suggests this does not have a large effect on our results.

We note that \citet{Nakatani_2018b} \citep[and also e.g.][]{Gorti_2004} use treat the X-ray heating by simply assuming $f_X$ to lie in the range $10-40$ per cent depending on the relative fractions of atomic and molecular gas, regardless of levels of ionisation.
By comparison to Fig. \ref{fig:columnE_eps}, we may expect this could have a significant effect on the ability of X-rays to heat a wind and may contribute (alongside their inclusion of molecular cooling, their very low X-ray to EUV luminosity ratio, and their X-ray spectrum dominated by a very soft excess) to the fact that they do not see a significant direct role for X-ray in launching a wind (note that conversely when they assume $100$ per cent of the photoelectron energy thermalises they do indeed find a more significant role for X-rays).

%Thus,  By lowering the temperatures and hence pressures below the base, molecular cooling reduces the momentum flux to the wind and ultimately the mass loss rate.

%Since we cannot therefore make a reasonable direct estimate of mass loss rates, if we instead assume that the local density scales with column density, then, we should expect $\dot{M} \propto N$. Thus, the reduced column heated by any X-rays in the face of molecular cooling would lead to lower mass loss rates.

%Overall therefore, we expect that by reducing the heatable column, deflating the disc and lowering the momentum flux, molecular cooling should decrease the mass loss rates.

\section{Conclusions}
We have explored the ability of different bands of radiation to drive a thermal wind by irradiating the density grid of \citetalias{Wang_2017} using \textsc{mocassin} in order to probe potential systematic differences between models of photoevaporative winds. Such systematic differences include the fundamental approach to radiative transfer, the cooling processes included and the nature of the irradiating spectrum
We have further used a simple toy model of thermal equilibrium to rationalise the results of these experiments.
Here we set out the key findings before summarising what it will take to accurately determine photoevaporative mass loss rates.

\begin{enumerate}
    \item The ability of X-rays to heat a higher column than the EUV and hence launch an X-ray driven wind is a strong function of frequency, which results from balancing the attenuation of lower frequencies against the larger column over which higher frequencies dissipate their energy. The most effective band - if one assumes the spectrum is effectively a delta function with a single frequency present - is $\sim500~\mathrm{eV}$ for typical cooling rates and luminosities; changing the representative frequency to such a value would result in an X-ray driven - rather than EUV-driven - wind with mass loss rates a few times higher. $1000~\mathrm{eV}$ X-rays as used by \citetalias{Wang_2017} are mostly unable to drive a wind (though may be marginally able to do so from restricted radii).
    \item Moreover, a realistic spectrum contains a range of X-ray energies each contributing to heating according to the shape of the spectrum. While the most representative band changes as a function of column density and radius, making it difficult to pick any single value to use for an X-ray bin even for a given spectrum, $1000~\mathrm{eV}$ is not a very representative energy anywhere or for any of the spectra considered.
    \item The relative ability of different X-ray energies to drive a wind is also dependent on the luminosity with higher X-ray luminosities resulting in harder energies becoming effective (at fixed EUV luminosity).
    \item Optical forbidden line radiation and Lyman $\alpha$ cooling mostly operate at higher temperatures than those at which material typically becomes unbound, so make relatively small differences to thermal balance where the wind is launched. Hence our results are relatively insensitive to the treatment of these cooling mechanisms which varies between previous works.
    \item However, the high $\sim3\times10^4~\mathrm{K}$ temperatures found in the wind of \citetalias{Wang_2017} are a result of missing cooling in their work, namely the complete absence of optical forbidden line radiation and the treatment of Lyman radiation as optically thick and non-escaping. 
    \item Adiabatic cooling is a modest contributor to thermal balance over most of the grid compared to emission line radiation once all such sources are accounted for. It may be most significant in regions of high acceleration such as in the low density column near the z axis. However, its significance would likely be lower in a cooler, denser X-ray heated wind at least for the $>10^{30}~\mathrm{erg~s^{-1}}$ luminosities considered here.
    \item Molecular cooling can be relevant near the temperatures at which material becomes unbound. This manifests in our simulations, which lack molecular cooling, as hotter temperatures below the base than found in \citetalias{Wang_2017}. If generous assumptions are made about molecular abundances, molecular cooling - particularly from water - could play an important role in reducing the maximum column heated by X-rays and further preventing hard frequencies from having sufficient heating effect to launch a wind. However it is likely somewhat more challenging for it to completely invert our conclusion that winds should be X-ray driven.
    %However, it certainly affects the underlying structure of the disc which determines both the flaring of the disc and degree to which it intercepts radiation, as well as the nature of material being fed to the wind base.
\end{enumerate}

With all this is mind, we argue that it is crucial for modelling of thermal winds to play close attention to the choice of irradiating spectrum. If too hard (or too soft) an X-ray band is used, its ability to heat a  column of material exceeding that heated by the EUV is diminished and an EUV-driven wind will result. Whereas, intermediate X-ray frequencies of a few $100~\mathrm{eV}$ - as are, realistically, present in the ionising spectra of T Tauri stars - should be able to launch an X-ray driven wind.
This is likely a key origin of conflicts between the X-ray driven photoevaporation models \citep[e.g.][]{Owen_2012,Picogna_2019} and EUV-driven models \citepalias{Wang_2017}.
However a more appropriate  choice of monochromatic frequency whose effect is equivalent to that of a realistic 
multi-frequency spectrum is hard to define as it varies with columnm, radius and X-ray luminosity; moreover the spectral shape can change as a function of stellar X-ray luminosity \citep{Preibisch_2005,Ercolano_2021}. Thus, it is not possible to capture the complete behaviour of X-ray heated winds too well with a single X-ray frequency.

Furthermore, the ability of radiation to launch a wind, particularly where harder frequencies are concerned, is dependent on the ability of photoionisation heating to overcome the local cooling. Thus, accurately establishing a complete set of cooling processes and coolant abundances relevant to conditions at the wind base will prove key to determining the exact range of frequencies which are able to overcome this and contribute to an X-ray driven wind.
Since X-ray driven models so far \citep{Owen_2010,Owen_2011,Owen_2012,Picogna_2019,Ercolano_2021,Picogna_2021} all neglect some sources of cooling, it is likely that the true mass loss rates are somewhat lower than derived from those works but may be less pessimistic than EUV-driven models would suggest.
%\textit{In future work we therefore intend to incorporate molecular cooling into our modelling to determine to what degree this reduces the ability of X-rays to launch a wind, and the consequences this may have for mass loss rates}.

Finally, we note that mass loss rates would be expected to scale with the X-ray luminosity \citep{Owen_2012,Picogna_2019} (which usually scales with stellar mass). 
In this work we have highlighted the importance of softer X-ray $< 1000~\mathrm{eV}$ and so the relevant X-ray luminosity is that emitted across the most effective frequencies. These luminosities are around a factor $2$ lower than the typically quoted values (which cover the $500-5000~\mathrm{ev}$ range); such effects should be born in mind when choosing X-ray luminosities for disc evolution modelling - as has been explored further by \citet{Ercolano_2021} - and recognising the correct scaling of these properties with stellar properties should prove crucial to population synthesis.

\section*{Acknowledgements}
%We thank the anonymous reviewer...
We are grateful to Lile Wang for sharing the density grids used to conduct our simulations. We thank the anonymous reviewer for a careful reading and constructive suggestions that helped strengthen our arguments, as well as Richard Booth, and James Owen's group, for useful discussions on this work.
ADS thanks the Science and Technology Facilities Council (STFC) for a Ph.D. studentship and CJC acknowledges support from the STFC consolidated grant ST/S000623/1.
This work has also been supported by the European Union's Horizon 2020 research and innovation programme under the Marie Sklodowska-Curie grant agreement No 823823 (DUSTBUSTERS).
BE was supported by the Deutsche Forschungsgemeinschaft (DFG, German Research Foundation) Research Unit `Transition discs' (FOR 2634/2, ER 685/8-2) and the Excellence Cluster ORIGINS of the German Research Foundation under Germany´s Excellence Strategy – EXC-2094 – 390783311.
This work was performed using resources provided by the Cambridge Service for Data Driven Discovery (CSD3) operated by the University of Cambridge Research Computing Service (www.csd3.cam.ac.uk), provided by Dell EMC and Intel using Tier-2 funding from the Engineering and Physical Sciences Research Council (capital grant EP/P020259/1), and DiRAC funding from the Science and Technology Facilities Council (www.dirac.ac.uk)

\section*{Data Availability}
%The observational datasets were derived from sources in the public domain using the VizieR catalogue access tool...

%The code from which model data were generated is available from...
X-ray enabled \textsc{mocassin} is available from \href{https://github.com/rwesson/mocassin_xray}{https://github.com/rwesson/mocassin\_xray}.
Temperature grids may be shared on reasonable request to the lead author.

%%%%%%%%%%%%%%%%%%%%%%%%%%%%%%%%%%%%%%%%%%%%%%%%%%



\bibliographystyle{mnras}
\bibliography{biblio} % if your bibtex file is called example.bib

%%%%%%%%%%%%%%%%%%%%%%%%%%%%%%%%%%%%%%%%%%%%%%%%%%

%%%%%%%%%%%%%%%%% APPENDICES %%%%%%%%%%%%%%%%%%%%%

%%%%%%%%%%%%%%%%%%%%%%%%%%%%%%%%%%%%%%%%%%%%%%%%%%

% Don't change these lines
\bsp	% typesetting comment
\label{lastpage}
\end{document}